\documentclass[prd,a4paper,preprintnumbers,nofootinbib,
showpacs, 
]{revtex4}
\usepackage{graphicx}
\usepackage{amsfonts}
\usepackage{amssymb}
\usepackage{slashed} 
\usepackage{multirow}
\usepackage[section]{placeins}
\usepackage{amsmath}
\usepackage{mathrsfs}
\usepackage{amsmath}
\def\eq#1{{eq.~(\ref{#1})}}
\def\eqs#1#2{{eqs.~(\ref{#1})--(\ref{#2})}}

\def\Im{\mbox{Im}\,}
\def\Re{\mbox{Re}\,}

\def\hbar{\hspace{0pt}\raisebox{1pt}{$-$} \hspace{-7pt} h}

\def\5{\overline 5}

\DeclareMathOperator\erf{erf}
\newcommand{\be}{\begin{equation}}
\newcommand{\ee}{\end{equation}}
\newcommand{\bea}{\begin{eqnarray}}
\newcommand{\eea}{\end{eqnarray}}

{
\usepackage{color}
\usepackage{hyperref}
\def\hhref#1{\href{http://arxiv.org/abs/#1}{#1}} 

\definecolor{oucrimsonred}{rgb}{0.6, 0.0, 0.0}
\definecolor{persianblue}{rgb}{0.11, 0.22, 0.73}
\definecolor{forestgreen}{rgb}{0.13,0.35,0.13}
 \hypersetup{colorlinks, citecolor=oucrimsonred, linkcolor=persianblue, urlcolor=oucrimsonred}
\begin{document}
\preprint{CERN-TH-2016-091}
\title[]{Telling the spin of the di-photon resonance}

\date{\today}
\author{Marco Fabbrichesi$^{\dag}$}
\author{Michele Pinamonti$^{^{\ddag\,\ast}}$}
\author{Alfredo Urbano$^{\circ}$}

\affiliation{$^{\dag}$INFN, Sezione di Trieste, via Valerio 2, 34136 Trieste, Italy}
\affiliation{$^{\ddag}$INFN, Sezione di Trieste, Gruppo collegato di Udine}
\affiliation{$^\ast$SISSA, via Bonomea 265, 34136 Trieste, Italy}
\affiliation{$^{\circ}$Theoretical Physics Department, CERN, Geneva, Switzerland}

\begin{abstract}
\noindent  We argue that the spin of the 750 GeV resonance
can be determined at the 99.7\% confidence level in the di-photon channel with as few as 10 fb$^{-1}$ of luminosity. 
This  result  is  true  if  the resonance is produced by gluon fusion (independently of the selection cuts) 
while an appropriate choice of selection cuts is needed if quark production is sub-dominantly present---which is the case of the Kaluza-Klein gravitational excitation under the hypothesis of a spin-2 resonance. A proportionally larger luminosity is required if the model for the spin-2 resonance includes a dominant production by quarks or  in the absence of an efficient  separation of the signal from the background.
\end{abstract}
\maketitle
%
\vskip1.5em

\section{Motivations and summary} 

The presence of a new state---a resonance in the di-photon channel in the first run of the LHC at 13 TeV~\cite{750}---will soon be either confirmed or disproved. Meanwhile,
the mere possibility of its existence behoves us to look, first and foremost, for the  identification of its other properties beside the mass. Only starting from such a full description---inclusive of mass, spin, parity and production and decay channels---it will be possible to begin to sift through the many possibilities of physics beyond the Standard Model (SM) that might account for its existence. 

In this paper, we focus on the determination of the spin  of such a new particle after its discovery. 
This problem has been discussed  for the Higgs boson by both theorists~\cite{higgs} and experimentalists~\cite{Aad:2013xqa}. It is therefore only a matter of adopting some of these analyses to the present case of a resonance with a mass of about 750 GeV. While a complete study must be based on the simultaneous use of different decays, we restrict ourselves to the di-photon channel because it is here that the resonance has been seen and will mostly likely be hunted down. In this channel, barring higher spin values, only two possibilities, spin-0 and 2, need to be discussed~\cite{Yang:1950rg}. 
 
 While the significance required for a discovery has been set very high at the $5\, \sigma$ level to avoid the risk of spurious results from even very unlikely background fluctuations, once the new particle discovery has been established, the determination of its spin can be accepted at the more mundane value of $3\,\sigma$ (99.7\% confidence level), as done, for example, for the Higgs boson itself~\cite{Aad:2013xqa}. 

In general, the discrimination between  different spin hypotheses improves if the background can be  subtracted in a reliable manner. We discuss  $_s\mathcal{P}lot$~\cite{Pivk:2004ty}---a procedure  that provides such a separation---and compare the significance of the spin determination for the signal alone and together with the un-subtracted background. 
 
We use and compare a log likelihood ratio (LLR) and a center-edge asymmetry~\cite{Dvergsnes:2004tw} to discriminate between the possible spin hypotheses. The main uncertainty arises from the systematic error in the definition of the spin-2 model because of the different  production mechanisms which have different angular distributions. If we assume that the spin-2 particle is almost completely produced by gluon fusion,  it will be possible not only to discovery the existence of the new state but also fix its spin with as few as 10 fb$^{-1}$ of luminosity in the di-photon channel. 
The number of events required for spin discrimination depends on the selection cuts utilised when the spin-2 particle is assumed to be significantly produced also from quarks. We discuss in detail the spin-2 model  embodied by the lightest  Kaluza-Klein graviton excitation after implementing the current constrains of its couplings to gluon, quark and photons. In this case an appropriated choice of selection cuts makes possible the spin identification again with only 10 fb$^{-1}$ of luminosity. A proportionally larger luminosity is required if the model for the spin-2 resonance includes a dominant production by quarks or  in the absence of an efficient  separation of the signal from the background.

We conclude by reviewing the potential relevance of terms of interference between signal and background~\cite{Dixon:2003yb} because they may play  a role more important for the 750 GeV resonance than in the Higgs boson case.

Two analyses of the 750 di-photon resonance recently appeared in which the issue of the spin determination is  discussed. The authors of ref.~\cite{Panico:2016ary}  utilise a framework  which cleverly exploits an encoding of the spin properties and production modes in the cross sections. It is an analysis that is different from ours  and that can be seen as complementary.  The authors of  ref.~\cite{Bernon:2016dow} discuss  how to characterise the new state by means of  distributions of a number of kinematical variables in various decay channels.

\section{Fit of the invariant mass distribution} \label{sec:Fit}
The first step in the analysis consists in extracting from the di-photon invariant mass $m_{\gamma\gamma}$ data  the value of the parameters of the models describing signal and background. 

We use the most recently published data from the ATLAS collaboration~\cite{ATLASMoriond} on the distribution of the di-photon mass invariant to fit the parameters of the signal and the background. While comparable data are also available from the CMS collaboration~\cite{CMS:2016owr}, we use those of the ATLAS collaboration because of the higher significance of the signal and their inclusion  of the  angular distributions in the published results. These experimental angular distributions are discussed in the appendix.

The signal is modelled by  a Breit-Wigner distribution 
\begin{equation}
f_{\rm BW}(m_{\gamma\gamma}) = \frac{2\mathcal{N}_{\rm BW}}{\pi} \left[
\frac{\Gamma_X^2 M_X^2}{\left(m_{\gamma\gamma}^2 - M_X^2\right)^2 + m_{\gamma\gamma}^4 \Gamma_X^2/M_X^2}\right]~.
\end{equation}

At this level, models of the resonance with different spin differ only by the overall normalisation of their respective signals and are not distinguishable. Fits of the invariant mass for different spins in the published data refer to differences in the selection cuts. We use the data after the Higgs-like cuts: $E_T^{\gamma 1} > 40 \, m_{\gamma\gamma}$ GeV, $E_T^{\gamma 2} > 30 \, m_{\gamma\gamma}$ GeV.

The background is described by  a family of nested functions 
with an increasing number of degrees of freedom
\begin{equation}
f_{(k)}(x;b,{a_k}) = \mathcal{N}_{\rm BG} \left(
1 - x^{1/3}
\right)^b
x^{\sum_{j = 0}^k a_j(\ln x)^j}~,
\end{equation}
with $x \equiv m_{\gamma\gamma}/\sqrt{s}$ and $b,\, a_j$ the parameters of the functions. 

\begin{figure}[!htb!]
\centering
  \includegraphics[width=.5\linewidth]{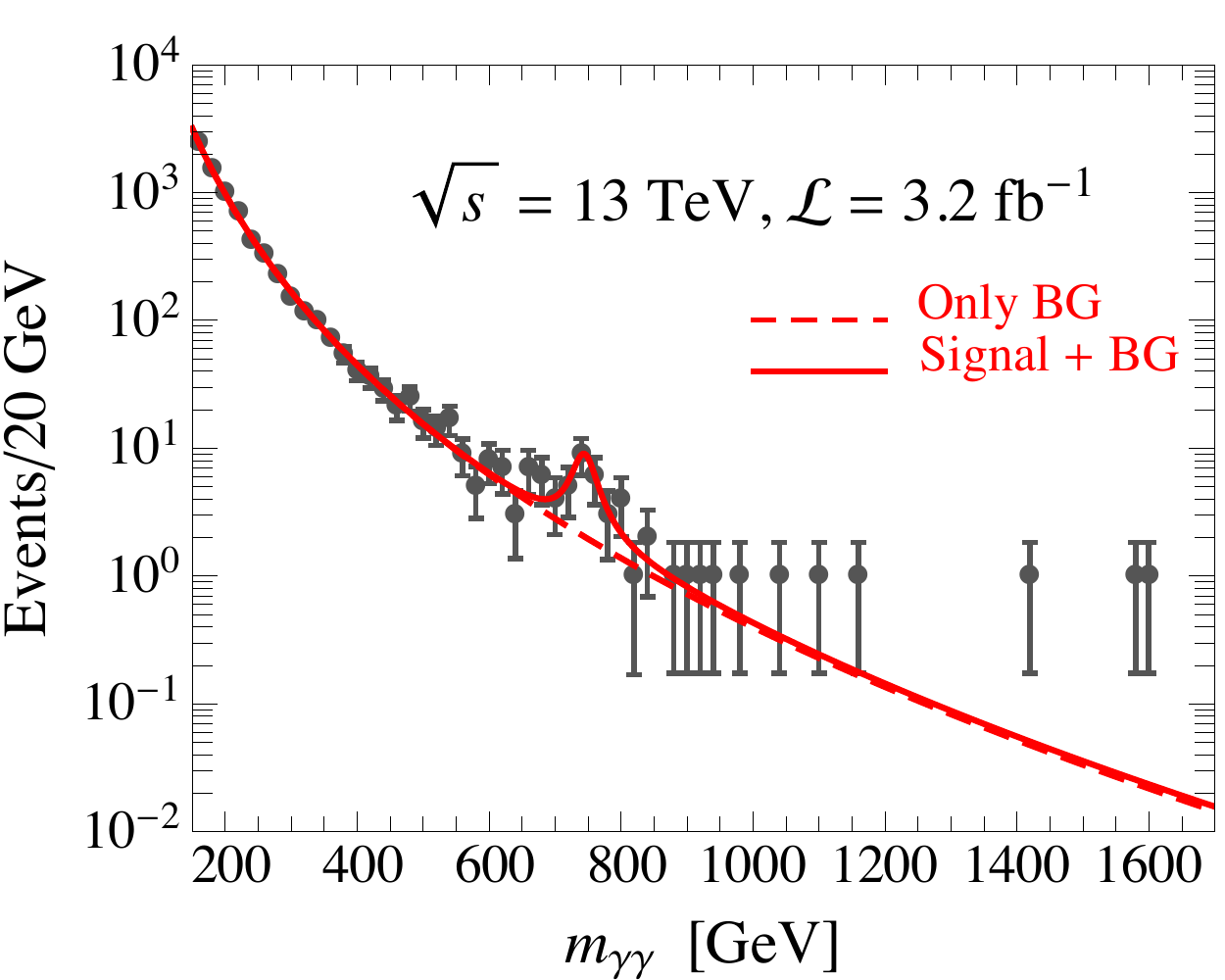}
\caption{\em 
The fit of the invariant mass distribution with selection cuts: $E_T^{\gamma 1} > 40 \, m_{\gamma\gamma}$ GeV, $E_T^{\gamma 2} > 30 \, m_{\gamma\gamma}$ GeV. Data digitally extracted from~\cite{ATLASMoriond}.
}
\label{fig:Fit}
\end{figure}

We perform a maximum likelihood (MLL) fit, the result of which is shown in fig.~\ref{fig:Fit}. It nicely agree with what reported by the experimental collaboration.
In table~\ref{tab:Fit} we collect the best-fitted values and errors thereof.
\begin{table}[htp]
\caption{{\it Result of the MLL fit in the invariant  mass range from 700 to 840 GeV  with selection cuts: $E_T^{\gamma 1} > 40 \, m_{\gamma\gamma}$ GeV, $E_T^{\gamma 2} > 30 \, m_{\gamma\gamma}$ GeV. The number of events refers to the signal-plus-background fit. }}
\begin{center}
\begin{tabular}{|c|c|c|c|c|c|c|}
\hline
\multirow{2}{*}{}  & \textbf{Normalisation} & \textbf{Mass [GeV]} & \textbf{Width [GeV]} & \textbf{Normalisation} & \textbf{BG coeff.} & \textbf{BG coeff.} \\
  & $\log_{10}(\mathcal{N}_{\rm BW})$ & $M_X$ & $\Gamma_X$ & $\log_{10}(\mathcal{N}_{\rm BG})$ & $a_0$ & $b$ \\
  \hline
 \textbf{Fit} & $1.04\pm 0.24$ & $745 \pm 8$ & $42\pm 27$ &  $-1.01 \pm 0.07$ & $-3.0 \pm 0.5$ &  $11.2 \pm 4.5$   \\
  \hline\hline
  & \multicolumn{3}{c|}{\textbf{Signal}} &  \multicolumn{3}{c|}{\textbf{Background}}   \\
  \hline
   \textbf{\# events }  &  \multicolumn{3}{c|}{$18.0 ^{+13.3}_{-7.6}\,~(\sigma_X = 5.6  ^{+4.2}_{-2.4}\, \mbox{fb}^{-1} )$ } &  \multicolumn{3}{c|}{$12.4^{+2.3}_{-1.9}$} \\
  \hline
\end{tabular}
\end{center}
\label{tab:Fit}
\end{table}%

A similar procedure can be followed in the case of the data after  looser cuts ($E_T^{\gamma 1}$ and $E_T^{\gamma 2}  > 55$ GeV).  We do not attempt such a fit here because it depends on a non-analytic modelling of the background, and simply rely on   the experimental collaboration that  gives  25 events for the signal and 45 for the background~\cite{ATLASMoriond}.

This MLL  fit provides us with an estimate of the  probability  that given model $B$ for the  background  and  $S$ for the signal, the data are in agreement with the expectations of  having  such a signal on top of the background, $P(data|S+B)$, as opposed to the background  alone, $P(data|B)$. Such an estimate is an example of  hypothesis testing and is usually  parametrised  in terms of a likelihood ratio ${\cal L}= -2 \log P(data|S+B)/P(data|B)$. 

The ATLAS collaboration cites  a significance of $\mathcal{Z}=3.9$ ($\mathcal{Z}=3.6$) for the fit in what they call ``spin-0 (spin-2) analysis" or, which is the same, the probability for the data to be a statistical fluctuation is, for both cases, of the order of or less than $10^{-3}$. The different significances for the two analyses are due to the higher background entering the selection in the case of the ``spin-2 analysis"---which contains more events in the forward direction where we also find most of the background.


\section{Theoretical inputs} 

Since we look at the di-photon channel, the spin of the resonance cannot be 1~\cite{Yang:1950rg}.\footnote{To avoid this theorem one must add an epicycle and assume that the putative spin 1 resonance decays into a photon and a scalar state which then decays into two almost collinear photons.} We therefore need only discuss the cases of spin-0 and 2.

\subsection{Spin and angular distributions}
\label{sec:Spin}

\begin{figure}[!ht]
\centering
  \includegraphics[width=.4\linewidth]{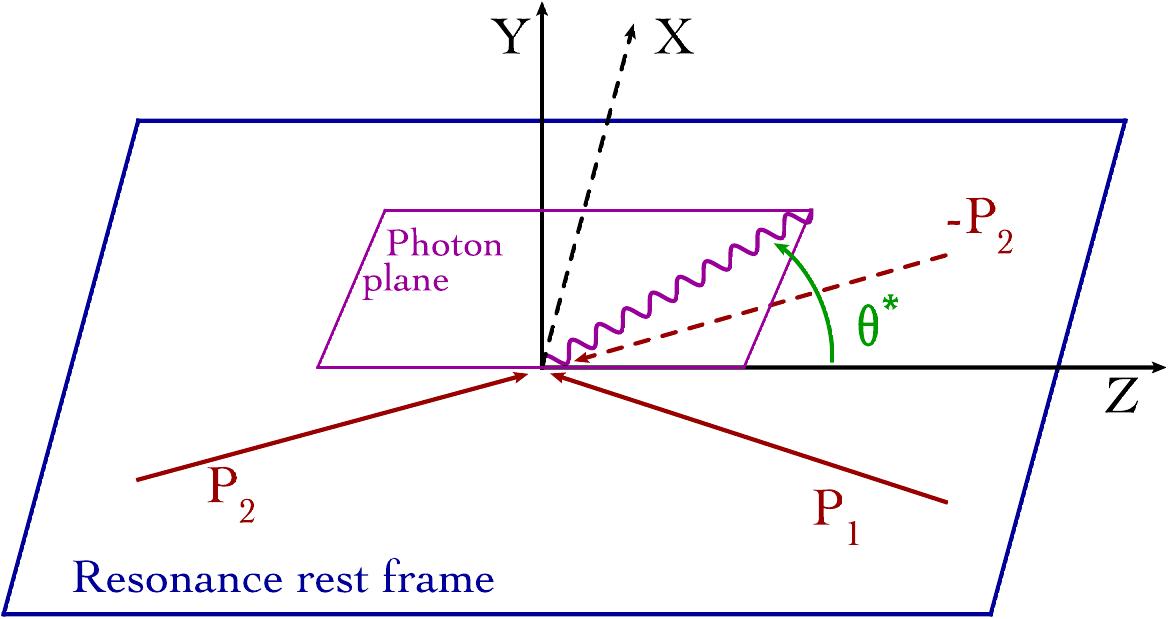}
\caption{\em 
Pictorial representation of the CS frame.
The CS frame  is constructed by boosting 
the event to the resonance's rest frame, and 
defining
the Z-axis as the inner bisector of the two partonic momenta $P_1$ and $-P_2$ (which are no longer collinear
in the resonance's rest frame). The X-axis becomes the outer bisector.
The photon scattering angle $\theta^*$ 
is measured w.r.t. the Z-axis.
}
\label{fig:CSFrame}
\end{figure}

Considering the di-photon decay $X_J \to \gamma\gamma$, informations about the spin $J$ of the decaying resonance  $X_J$ 
can be extracted from the distribution of the 
photon scattering angle in the Collins-Soper (CS) frame~\cite{Collins:1977iv} (see fig.~\ref{fig:CSFrame}).

For a spin-0 resonance, we have 
\begin{equation}\label{eq:AngoloSpin0}
\frac{1}{N^{\rm spin-0}}\times \frac{dN^{\rm spin-0}}{dz} = \frac{1}{2}~,
\end{equation}
with $z \equiv \cos\theta^*$ the cosine of CS scattering angle defined as
\begin{equation}\label{eq:CSdefinition}
\cos\theta^* = \frac{\sinh\Delta\eta_{\gamma\gamma}}{\sqrt{1+\left(\frac{p_T^{\gamma\gamma}}{m_{\gamma\gamma}}\right)^2}}
\frac{2p_T^{\gamma_1}p_T^{\gamma_2}}{m_{\gamma\gamma}^2}~.
\end{equation}
The photon pair transverse momentum is $p_T^{\gamma\gamma} = \sqrt{(p_T^{\gamma_1})^2+(p_T^{\gamma_2})^2 
+2p_T^{\gamma_1}p_T^{\gamma_2}\cos\Delta\varphi_{\gamma\gamma}}$, with $\Delta\varphi_{\gamma\gamma}$ the azimuthal angle between the two photons;
$\Delta\eta_{\gamma\gamma}$ is the pseudo-rapidity difference
$\eta_{\gamma_1} - \eta_{\gamma_{2}}$.

For a spin-2 resonance the production mode affects the
angular distribution of the final state photons and we have two possibilities:
\begin{eqnarray}
\frac{1}{N_{gg}^{\rm spin-2}}\times \frac{dN_{gg}^{\rm spin-2}}{dz} &=& \frac{5}{32}\left(
1+ 6z^2 + z^4
\right)~,\label{eq:Spin2gg}\\
\frac{1}{N_{qq}^{\rm spin-2}}\times \frac{dN_{qq}^{\rm spin-2}}{dz} &=& 
\frac{5}{8}\left(
1 - z^4
\right)~.\label{eq:Spin2qq}
\end{eqnarray}
In the presence of both production mechanisms the expected angular distribution is
\begin{equation}
\frac{1}{N^{\rm spin-2}}\times \frac{dN^{\rm spin-2}}{dz} = 
\frac{5}{32}\left(
1+ 6z^2 + z^4
\right)(1-f_{qq}) + \frac{5}{8}\left(
1 - z^4
\right)f_{qq}~,
\end{equation}
where $f_{qq} \equiv N_{qq}^{\rm spin-2}/N^{\rm spin-2}$ is the relative weight of the production by quarks.

The two \eqs{eq:Spin2gg}{eq:Spin2qq} show the main problem with any model of a spin-2  resonance: while  the gluon production has an angular distribution that is clearly distinguishable from that of spin-0, the case of quark production has an angular distribution  rather similar to the spin-0 case and therefore the separation between the two models is more difficult.

The angular distributions in eqs.~(\ref{eq:AngoloSpin0},\,\ref{eq:Spin2gg},\,\ref{eq:Spin2qq})
are strictly valid only considering parton-level events without  showering and detector simulation.
They are considerably distorted once detector acceptances are included.
It is therefore important to investigate the angular distributions 
in a more complete framework. 

The total cross section for the resonant process $pp\to X_{J} \to \gamma\gamma$ is given, in full generality, by 
\begin{equation}
\sigma(pp\to X_J \to \gamma\gamma) = \frac{2J + 1}{M_{X_J}\Gamma_{X_J} s}
\left[
\sum_{\mathcal{P}}C_{\mathcal{PP}}\Gamma(X_{J} \to \mathcal{PP})
\right]\Gamma(X_{J} \to \gamma\gamma)~,
\end{equation}
$C_{\mathcal{PP}}$ 
are the dimensionless  partonic
integrals whose numerical values can be found, for instance, in~\cite{Franceschini:2015kwy}.
Production mechanism and decay modes depends on the 
microscopic interactions of the resonance.

We do not focus on any particular model
and do not attempt to construct the most general effective field theory of the resonance $X$. 
In this paper, we only look at those effective interactions that are relevant
to determine the spin of the resonance.

\subsection{Spin-0}

\begin{figure}[!htb!]
\centering
\minipage{0.5\textwidth}
  \includegraphics[width=0.8\linewidth]{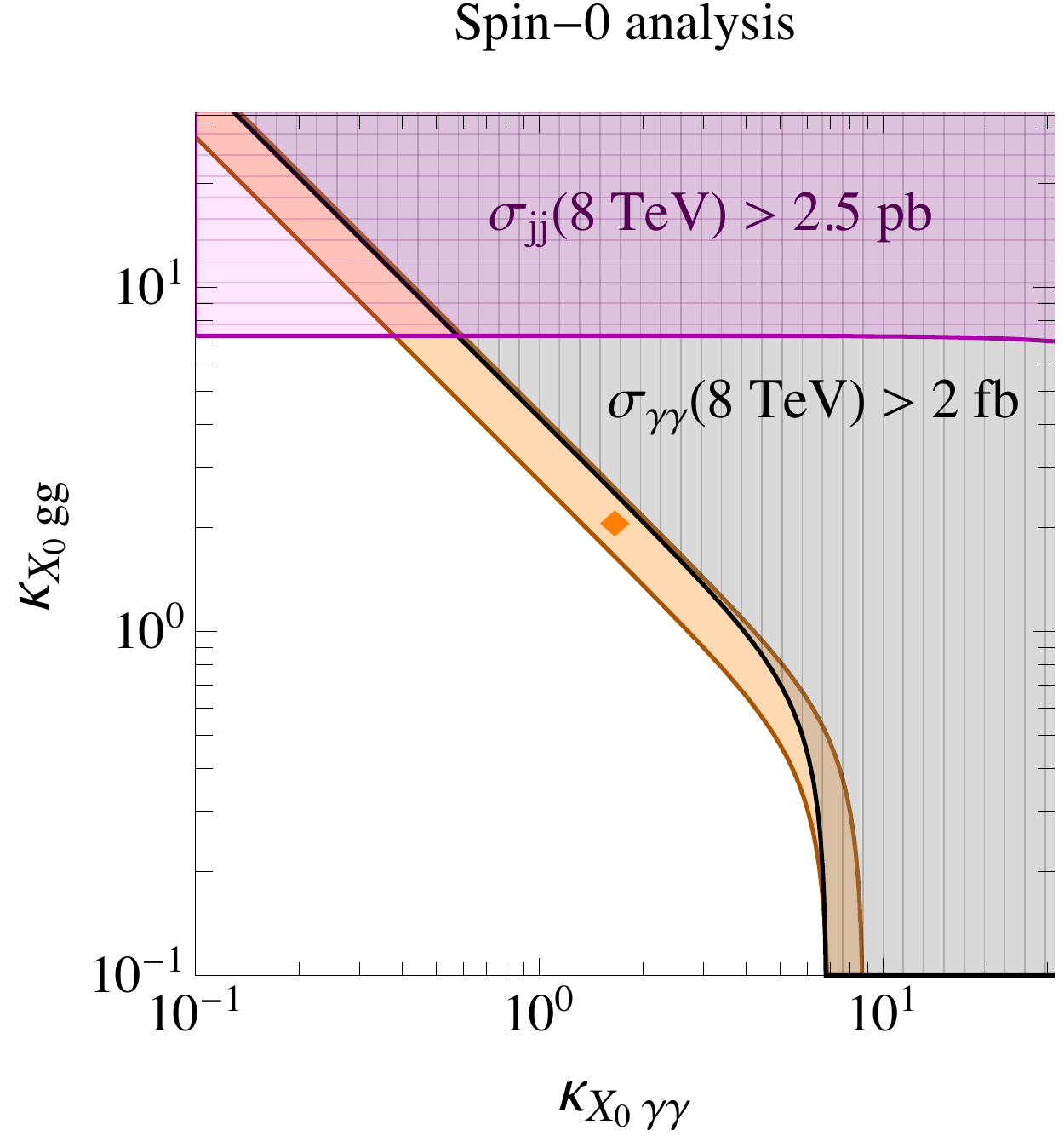}
\endminipage\hfill
\minipage{0.5\textwidth}
  \includegraphics[width=0.8\linewidth]{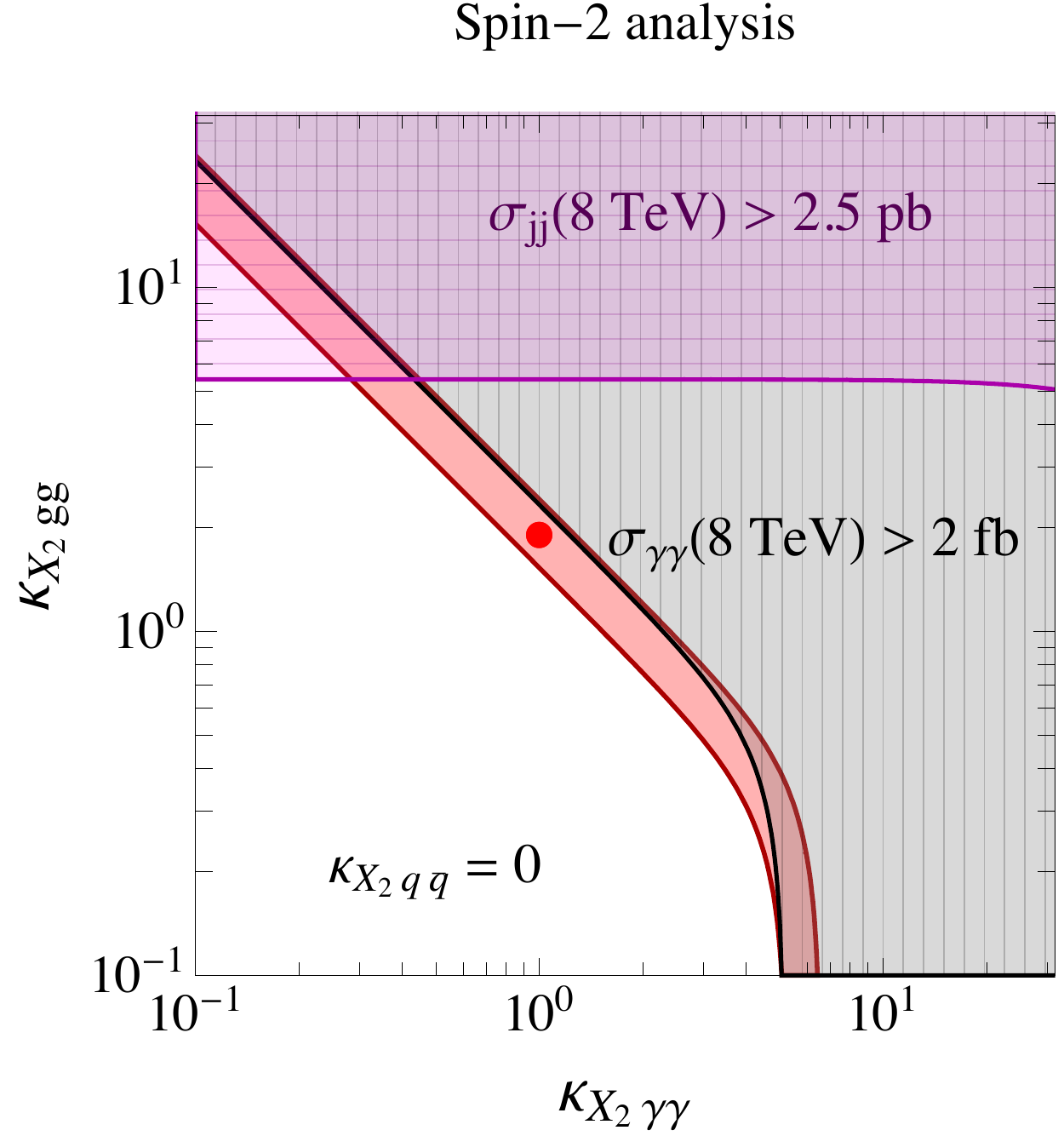}
\endminipage
\caption{\em Constraints on the parameters from data at 8 TeV~\cite{Franceschini:2015kwy}. 
The left (right) panel refers to the case of a spin-0 (spin-2) resonance.
The regions shaded in orange and red 
correspond to $\sigma(pp\to X_J \to \gamma\gamma) = [4-10]$ fb at $\sqrt{s} = 13$ TeV.
The orange diamond and red circle 
correspond to the specific values of the couplings used in section~\ref{sec:Methods}.
}
\label{fig:ParameterSpace}
\end{figure}

We assume the following effective interaction  Lagrangian
\begin{equation}\label{eq:LagSpin0}
\mathcal{L}_{\rm spin-0} =  -\frac{1}{4\Lambda}\left(
\kappa_{X_0 WW} W_{\mu\nu}^{A}W^{A\,\mu\nu}  + \kappa_{X_0 \gamma \gamma} A_{\mu\nu}A^{\mu\nu}
+  \kappa_{X_0 gg} G^a_{\mu\nu}G^{a\,\mu\nu}
\right)~,
\end{equation}
where $A_{\mu\nu}$, $W_{\mu\nu}^A$, and  $G_{\mu\nu}^a$ are
the field strength tensors for the SM gauge groups $U(1)_Q$, $SU(2)_L$, and $SU(3)_C$, respectively.  
We do not introduce any interactions with SM quarks, 
since the angular distribution in eq.~(\ref{eq:AngoloSpin0}) does not depend on the detail of the production mechanism.
From eq.~(\ref{eq:LagSpin0}) we extract the following decay widths
\begin{equation}
\Gamma(X_0 \to \gamma\gamma) = \frac{\kappa_{X_0\gamma\gamma}^2 M_{X_0}^3}{64\pi \Lambda^2}~,~~~~~~~
\Gamma(X_0 \to gg) = \frac{\kappa_{X_0 gg}^2 M_{X_0}^3}{8\pi \Lambda^2}\, .
\end{equation} 
We show the relevant parameter space in the left panel of fig.~\ref{fig:ParameterSpace} in which we fixed $\Lambda = 10$ TeV.
The regions shaded in orange 
corresponds to $\sigma(pp\to X_0 \to \gamma\gamma) = [4-10]$ fb at $\sqrt{s} = 13$ TeV, 
that is the value needed to fit the observed excess.
According to the result of the fit in section~\ref{sec:Fit}, we set the total decay width $\Gamma_{X_0} = 42$ GeV and we take $M_{X_0} = 745$ GeV for the mass.
We also show in the left panel of fig.~\ref{fig:ParameterSpace} 
the bounds from data at $\sqrt{s} = 8$ TeV. The most relevant constraints come from 
di-jet and di-photon searches~\cite{Franceschini:2015kwy,Knapen:2015dap}.

\subsection{Spin-2}

We assume the following interaction effective Lagrangian
\begin{equation}\label{eq:LagSpin2}
\mathcal{L}_{\rm spin-2} = -\frac{1}{\Lambda}\sum_{f = q,l}\kappa_{X_2 f\bar{f}} T_{\mu\nu}^f X_2^{\mu\nu} 
 -\frac{1}{\Lambda}\sum_{V = Z,W,\gamma, g}\kappa_{X_2 VV} T_{\mu\nu}^V X_2^{\mu\nu}~,
\end{equation}
where the explicit expression for the various components of the energy-momentum tensor can be found in~\cite{Giudice:1998ck}.
In full generality we 
keep separate coupling parameters $\kappa_{X_2 f\bar{f}}$ and $\kappa_{X_2 VV}$ with SM fermions and gauge bosons.
In the minimal Randall-Sundrum  scenario the couplings of the spin-2 particle---identified with the lightest Kaluza-Klein (KK) graviton excitation---are universal~\cite{Randall:1999ee}.

The case of the KK graviton will be explored in section~\ref{sec:KKGraviton}.
In order to better illustrate our methodology, and facilitate the comparison with the spin-0 case, 
we start our discussion from a very simple phenomenological toy model in which 
all the interactions in eq.~(\ref{eq:LagSpin2}) but the couplings with photons and gluons are set to zero.
 In this setup the relevant decay widths are the following
\begin{equation}
\Gamma(X_2 \to \gamma\gamma) = \frac{\kappa_{X_2 \gamma\gamma}^2 M_{X_2}^3}{80\pi \Lambda^2}~,~~~~~
\Gamma(X_2 \to gg) = \frac{\kappa_{X_2 gg}^2 M_{X_2}^3}{10\pi \Lambda^2}~.
\end{equation}
In this case we expect to reproduce the angular distribution given in eq.~(\ref{eq:Spin2gg}), and the differences between spin-0 and spin-2 
are therefore maximised.

We show the relevant parameter space in the right panel of fig.~\ref{fig:ParameterSpace}.
As before, we impose for the total decay width $\Gamma_{X_2} = 42$ GeV and we take $M_{X_2} = 745$ GeV.

The case of a narrow width is not excluded by current data.
A resonance with mass $M_X$ and narrow width $\Gamma_X$, in fact, inherits---at the level of detector simulation---an effective width comparable with the energy resolution of the detector
at invariant mass $m_{\gamma\gamma} \approx M_X$, that is roughly $8$ GeV at $m_{\gamma\gamma} \approx 750$ GeV.
As we shall discuss in section~\ref{sec:KKGraviton}, the KK graviton falls into this class of models.


\section{Methods} \label{sec:Methods}

\subsection{Background and signal}

\begin{figure}[!htb!]
\centering
  \includegraphics[width=.6\linewidth]{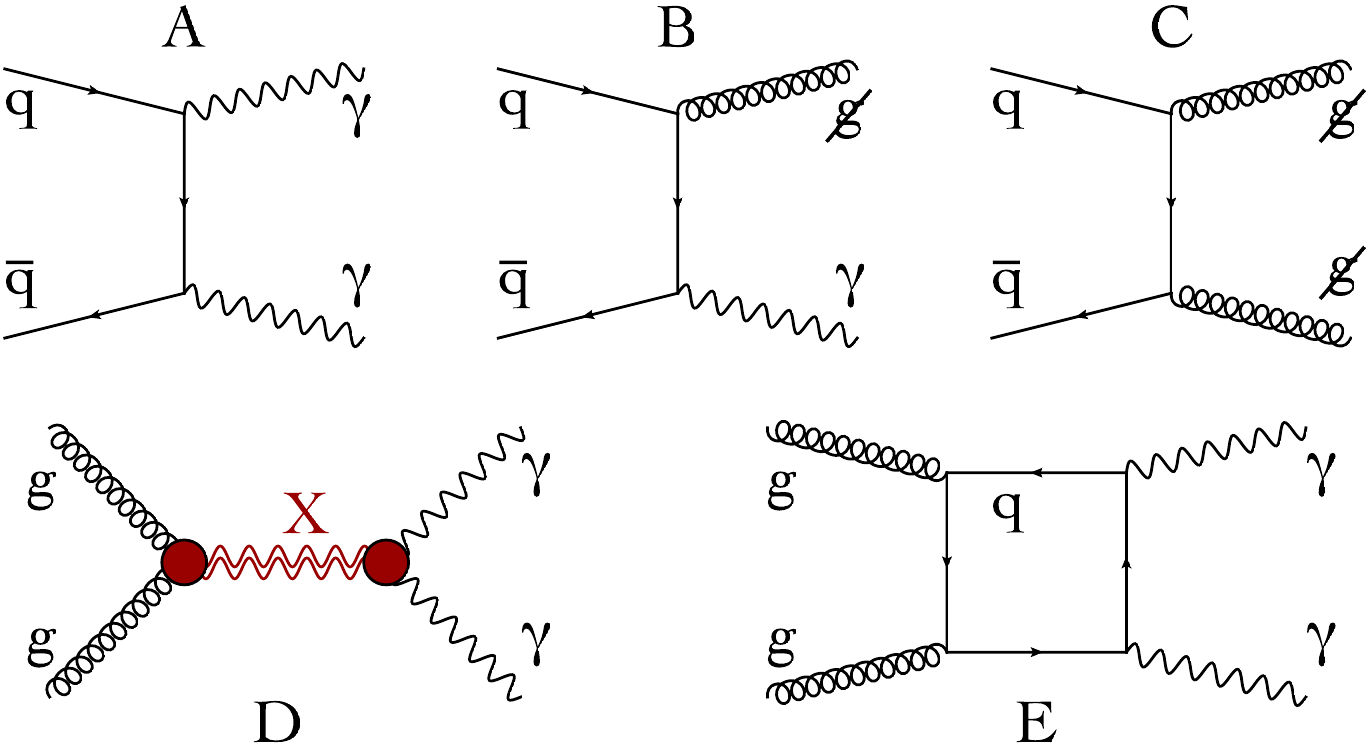}
\caption{\em 
Representative Feynman diagrams for background and signal.
}
\label{fig:Diagrams}
\end{figure}

The irreducible background is comprised of tree-level non-resonant di-photon quark annihilation (diagram $A$ in fig.~\ref{fig:Diagrams}).
Due to the $t$- and $u$-channel exchange of light quarks, this process is peaked in the forward direction in the relevant invariant mass range. This is not true at lower invariant masses---like those for  the analysis of the Higgs boson---where the background shows a different angular distribution.

One-loop gluon fusion into di-photon final state (diagram $E$ in fig.~\ref{fig:Diagrams}) also contributes to the irreducible background.
The gluon fusion process competes with the tree-level quark annihilation for photon pairs of invariant mass less than $200$ GeV~\cite{Dicus:1987fk}
where the loop suppression is compensated by (in order of importance) the large gluon luminosity,
the accidentally large value of the scattering amplitude, and the coherent sum of all quark flavors in the loop.
In the invariant mass range $m_{\gamma\gamma} = [700-840]$ GeV, on the contrary, 
the one-loop gluon fusion process turns out to be suppressed with respect to the tree-level quark annihilation, and therefore largely subdominant for our purposes. 
We come back to these loop diagrams   in the appendix~\ref{app:A}.

The reducible background is mostly comprised of tree-level quark annihilation into one photon and one gluon (diagram $B$ in fig.~\ref{fig:Diagrams}) and two gluons (diagram $C$ in fig.~\ref{fig:Diagrams})  in the final state. These processes mimic the di-photon final state because of possible misidentification of gluons at the detector level.

In our simulations we focus only on the irreducible $q\bar{q}\to \gamma\gamma$ background. 
As shown in detail in~\cite{ATLASMoriond} the reducible background component generated from the probability for a jet to fake a photon 
is subdominant with respect to  the genuine di-photon pair production 
in the whole invariant mass range $m_{\gamma\gamma} = [200-2000]$ GeV.

The signal  is generated by $s$-channel resonant production of $X$ (diagram $D$ in fig.~\ref{fig:Diagrams}, in the case of spin-2 resonance produced via gluon fusion)
with subsequent di-photon decay.
We implement the Lagrangian in eqs.~(\ref{eq:LagSpin0},\ref{eq:LagSpin2}) in \texttt{FeynRules}~\cite{Alloul:2013bka}
following the benchmark example of the Higgs characterisation model~\cite{Artoisenet:2013puc}.

We generate background, signal, and background-plus-signal samples 
by means of the matrix-element plus parton-shower merging procedure.
We use \texttt{MadGraph5\_aMC@NLO}~\cite{Alwall:2014hca} (MG5 hereafter) supplemented by \texttt{Pythia\,6}~\cite{Sjostrand:2006za} for showering and \texttt{Delphes}~\cite{deFavereau:2013fsa} for detector simulations.
We generate signal samples via $pp\to X_{0,2} \to \gamma\gamma$, and we add processes with additional $1$ and $2$ jets
 matching the correct parton multiplicity using the MLM algorithm.
The merging separation parameter is set to $Q_{\rm cut} = 200$ GeV.
Irreducible background as well as signal-plus-background samples are generated following the same procedure.

\subsection{Selection cuts}

Only events within the interval $m_{\gamma\gamma}=$ [700-840] GeV are considered. A first cut is enforced on the rapidity and, following~\cite{ATLASMoriond},  the region $|\eta| > 2.37$ as well as the window  $\eta = [1.37-1.52]$ are excluded. 

Mainly for historical reasons, the analysis of the spin has been presented by the experimental collaborations with two different selection cuts in the transverse energy. In a perhaps misleading labelling, they have been referred to as ``spin-0'' and ``spin-2 analysis''. We rename them. The first one (\textit{tight cuts}) are like those used for the study of the Higgs boson. The other set (\textit{loose cuts}) allow for lower values of $p_T$ and populate the forward region. 

 The two sets  are \begin{equation}\label{eq:HiggsCuts}
\mbox{\underline{tight selection cuts}:}\quad (E_T^{\gamma_1} > 0.4~m_{\gamma\gamma}, ~E_T^{\gamma_2} > 0.3~m_{\gamma\gamma}) \, ,
\end{equation}
and
\begin{equation}\label{eq:LooseCuts}
\mbox{\underline{loose selection cuts}:}\quad (E_T^{\gamma_1} > 55~{\rm GeV},~E_T^{\gamma_2} > 55~{\rm GeV})    \, .
\end{equation}

The tight cuts remove most of the forward region, a region where  acceptance effects are important. Moreover, having the spin-0 case in mind,  these cuts were devised to reduce the contribution of the the background which is mostly in this region.
Notice that the tight cuts have the additional nice feature of reducing the correlation between the invariant mass and the angular distribution. On the other hand, the loose cuts were thought with the case of a spin-2 state in mind because of its angular distribution that, at least in the case of gluon production, peaks in the forward direction.

 We consider both selection cuts and compare their effectiveness.

\subsection{Unfolding the data with $_s\mathcal{P}lot$} 
\label{sec:sPlot}

In our analysis we assume that an effective way of separating the signal from the background has been implemented. This is a complicated problem with many subtle implications and different solutions. We illustrate this crucial step by means of
 $_s\mathcal{P}lot$~\cite{Pivk:2004ty}, a technique that makes it possible to unfold the data and isolate the signal.
This technique allows to reconstruct the distribution of  a \textit{control} variable by means of our knowledge of the \textit{discriminating} variable distribution. It is useful in those cases in which we know how to separate the signal from the background for the discriminating variable better than for the control variable.
 
\begin{figure}[!ht]
\centering
\minipage{0.5\textwidth}
  \includegraphics[width=.95\linewidth]{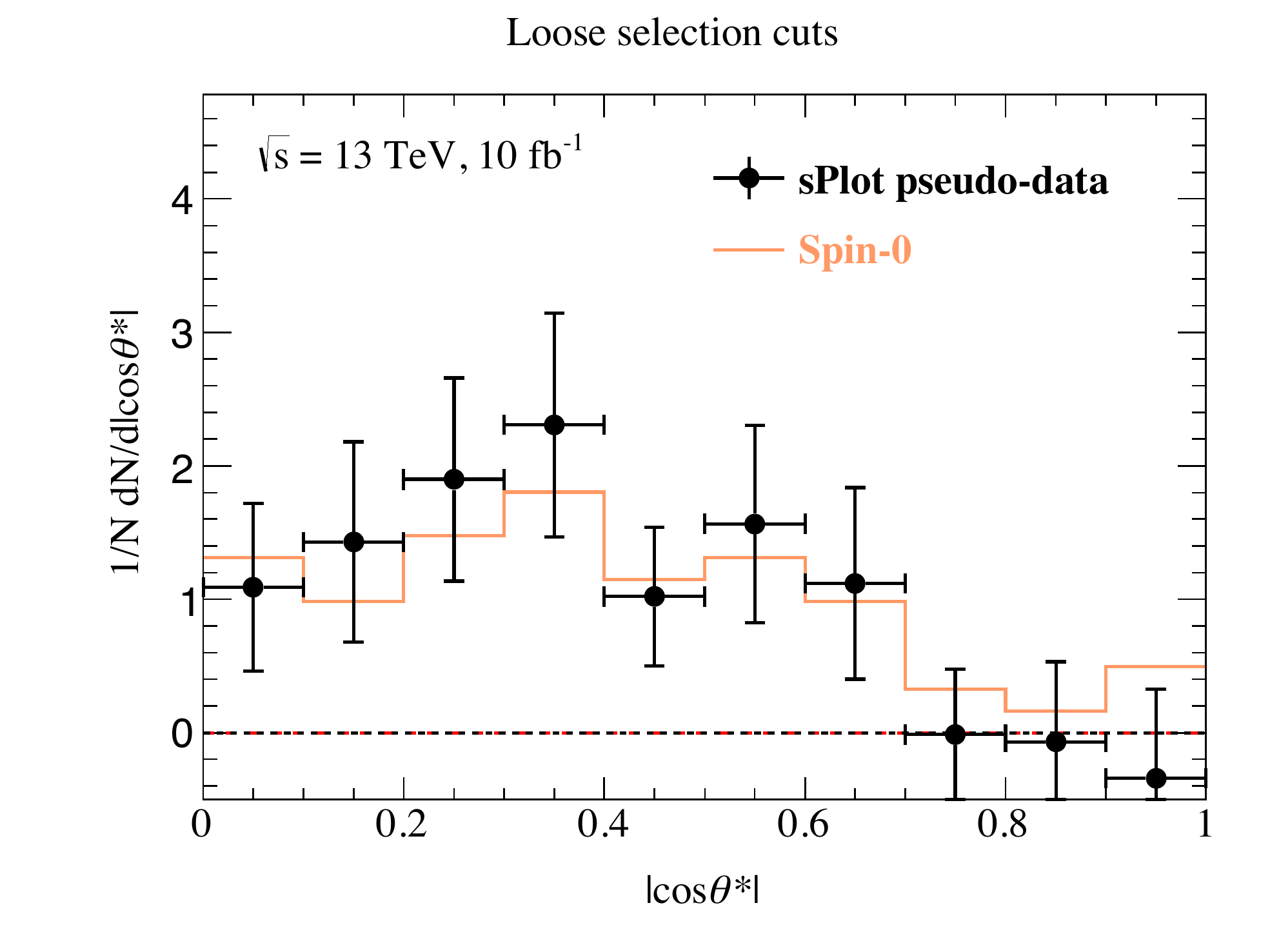}
\endminipage\hfill
\minipage{0.5\textwidth}
  \includegraphics[width=.95\linewidth]{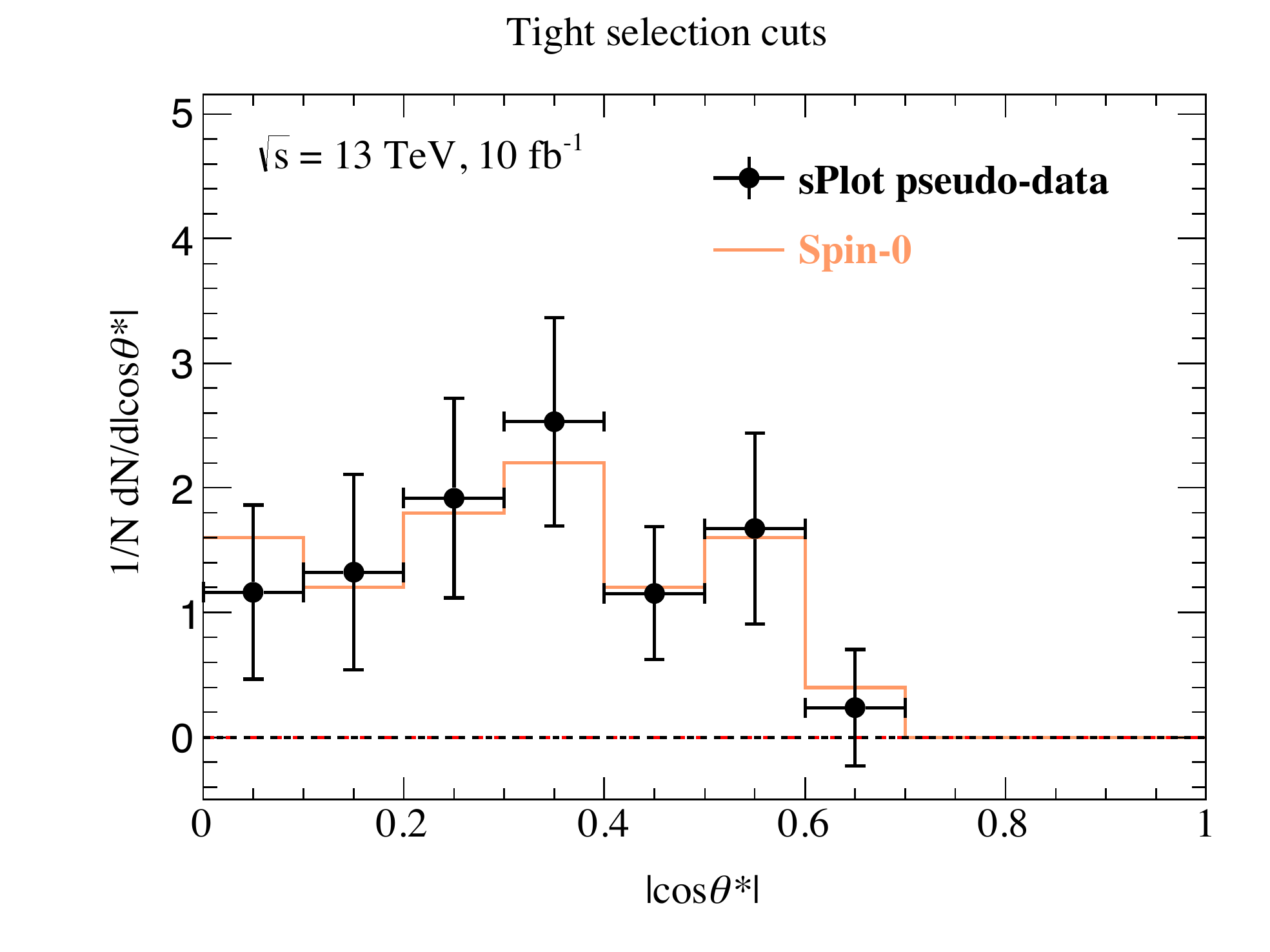}
\endminipage
\caption{\em Pseudo-data of angular distribution of the signal after unfolding by means of $_s\mathcal{P}lot$ compared with the Monte Carlo generated spin-0 signals for a 10 fb$^{-1}$ luminosity.
}
\label{fig:splot0}
\end{figure}\begin{figure}[!h]
\centering
\minipage{0.5\textwidth}
  \includegraphics[width=.95\linewidth]{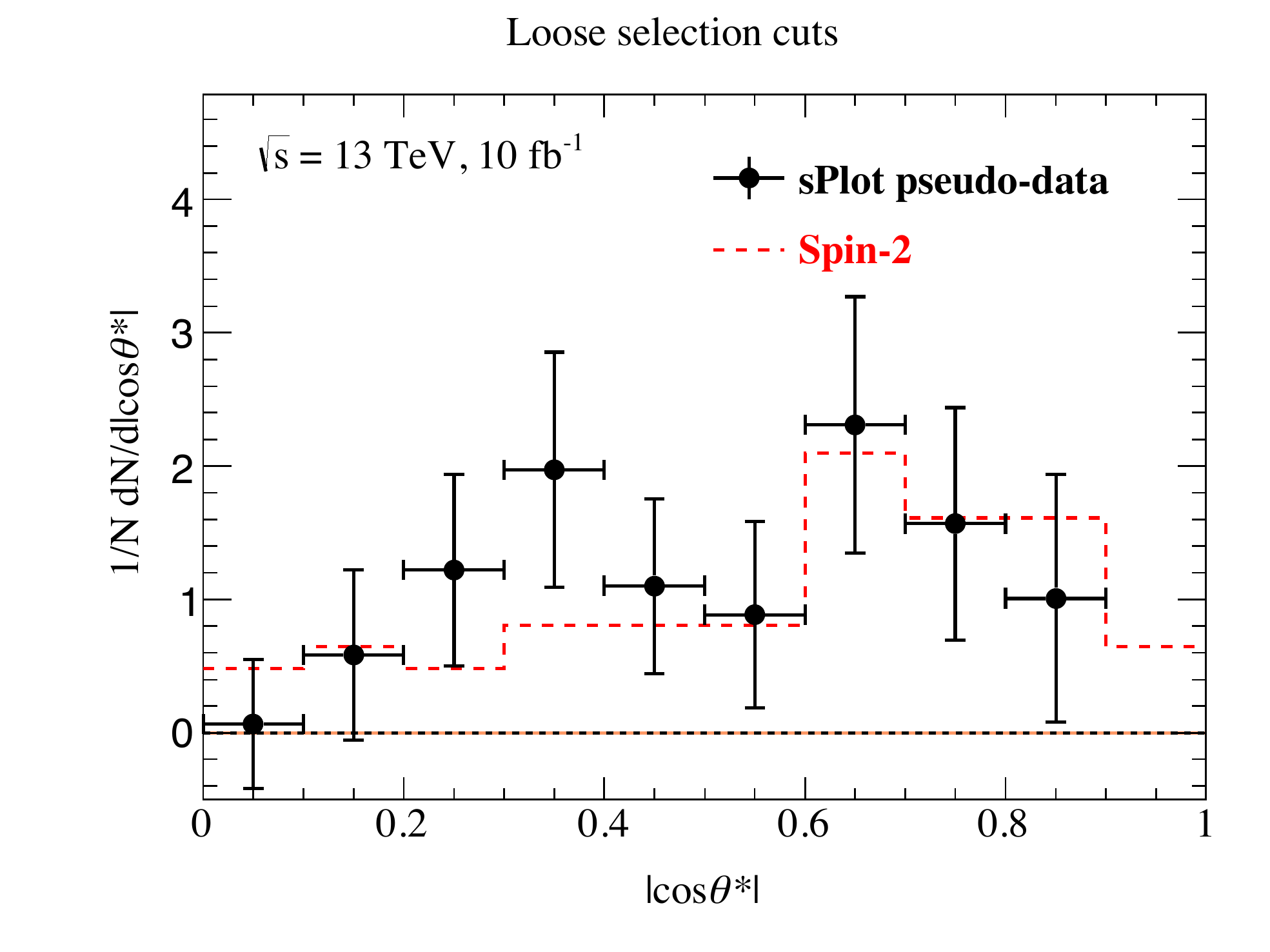}
\endminipage\hfill
\minipage{0.5\textwidth}
  \includegraphics[width=.95\linewidth]{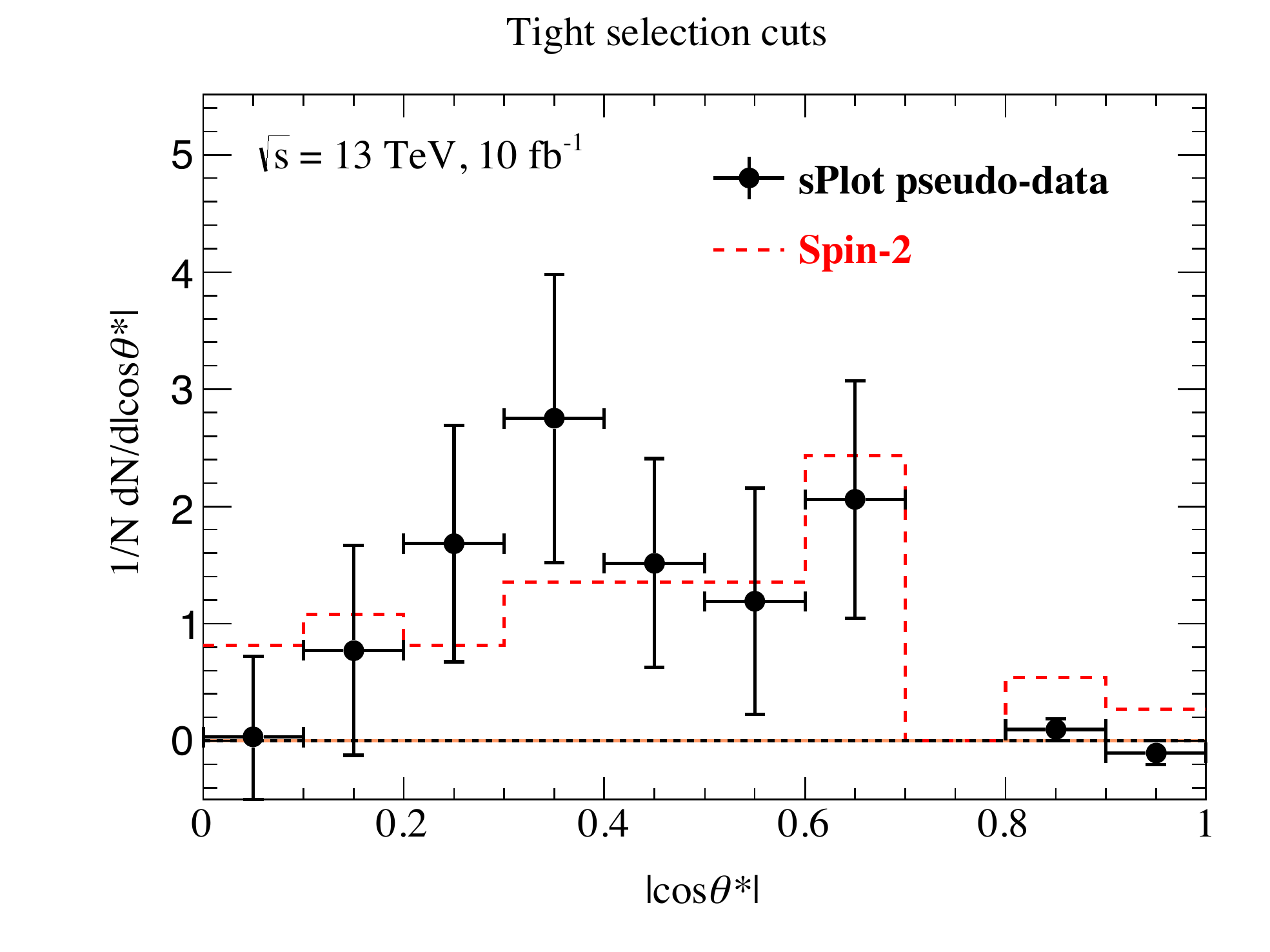}
\endminipage
\caption{\em Pseudo-data of angular distribution of the signal after unfolding by means of $_s\mathcal{P}lot$  compared with the the Monte Carlo generated spin-2 signals for a 10 fb$^{-1}$ luminosity.
}
\label{fig:splot2}
\end{figure}

The technique can be summarised as follows. Given a likelihood for $N_e$ events in which the events for the signal $N_S$ and the background $N_B$ are mixed:
\be
{\cal L} = \sum_{k=1}^{N_e} \ln \left[ N_S \times {\rm pdf}_S(x_k) + N_B \times {\rm pdf}_B(x_k) \right] - N_S - N_B \, ,
\ee
 one  computes first the {\it sWeights}
\be
_sW_n(x_k) = \frac{{\bf V}_{n}^S \times {\rm pdf}_S(x_k) + {\bf V}_{n}^B \times {\rm pdf}_B(x_k)}{\hat N_S \times {\rm pdf}_S(x_k) + \hat N_B \times {\rm pdf}_B (x_k)} \, ,
\ee
where $\hat N_{S,B}$ maximises the  likelihood $\cal L$ and $({\bf V}^{S,B}_{n})^{-1} = - \partial^2 {\cal L}/\partial N_n \partial N_{S,B}$. The discriminating variable $x_k$ is in our case $m_{\gamma \gamma}$, the di-photon invariant mass. 

The histograms for the control variable $y_k$---in our case  $\cos \theta$ in the bin $k$---are then generated by computing
\be
\bar{y}_{n} = \sum_{|y_{k}- \bar{y}_{n}| < \delta_{y}} {_s}W_{n}(x_{k}) \, ,
\ee
where $\delta_y$ and $\bar y_k$ are, respectively, the width and central value of the control variable in the given bin. The histogram constructed from this prescription is called the 
{\it sPlot} and provides us with the angular distribution of the signal independently of the background events.

The method works best if control and discriminating variables are uncorrelated---as they mostly are in the case of invariant mass and angular variables, with some dependence, as already discussed, on the selection cuts. 

$_s\mathcal{P}lot$ has been implemented within \texttt{Root}~\cite{root} and its use already championed in the case of the Higgs boson~\cite{higgs}. Figs.~\ref{fig:splot0} and \ref{fig:splot2} show the pseudo-data of angular distribution of the signal after unfolding by means of $_s\mathcal{P}lot$ compared with the Monte Carlo generated spin-0  and spin-2 signals for a 10 fb$^{-1}$ luminosity. The error bars and the reliability of the technique improve for higher luminosities. 

\section{Results} \label{sec:Results}

After generating background and signals, we can construct probability density functions (pdf)
for both angular and mass invariant distributions. The two spin hypotheses can then be discriminated by measuring either a LLR or an asymmetry on  randomly generating events weighted  by the angular pdf.

\subsection{Probability density functions}
\label{sec:PDFs}

\begin{figure}[!ht]
\begin{center}
\fbox{Tight selection cuts}
\end{center}
\vspace{-0.01cm}
\centering
\minipage{0.5\textwidth}
  \includegraphics[width=.8\linewidth]{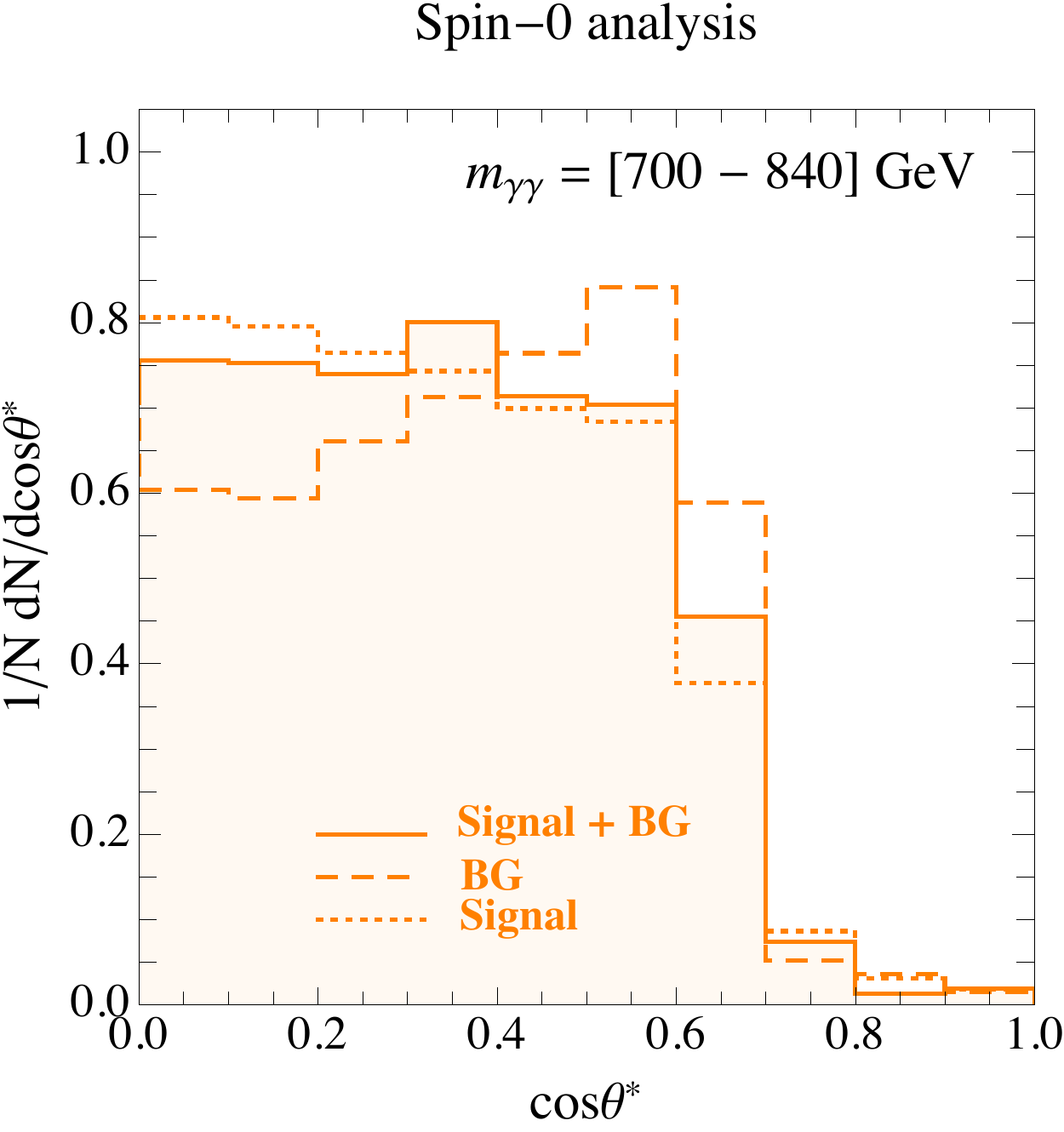}
\endminipage\hfill
\minipage{0.5\textwidth}
  \includegraphics[width=.8\linewidth]{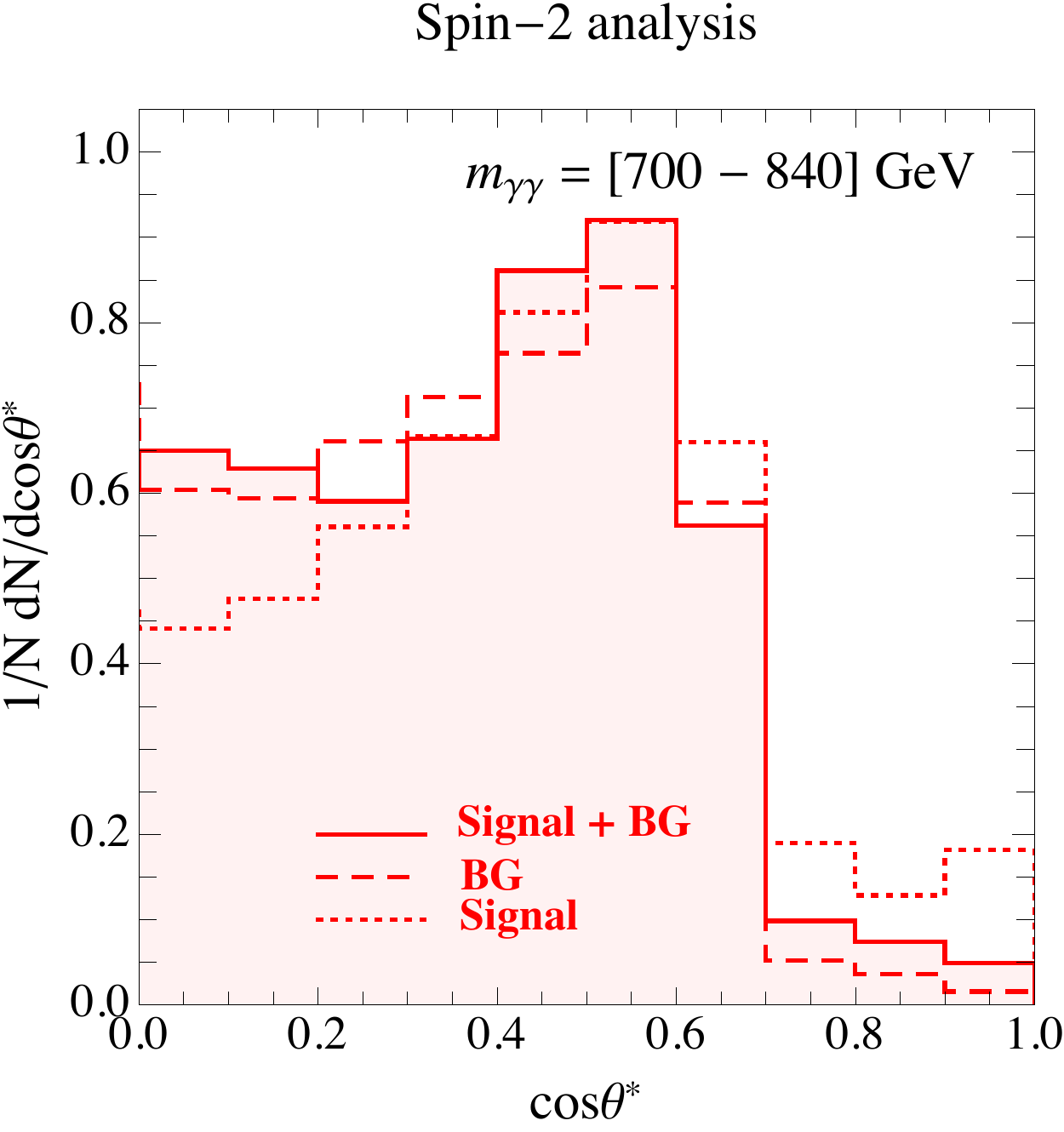}
\endminipage
\caption{\em 
Angular distributions (normalised to $1$ in the interval $\cos\theta^* \in [-1,1]$) for the tight set of selection cuts 
for the spin-0 (left panel, orange) and spin-2 (right panel, red) analysis.
}
\label{fig:Histograms}
\end{figure}

\begin{figure}[!ht]
\begin{center}
\fbox{Loose selection cuts}
\end{center}
\vspace{-0.01cm}
\centering
\minipage{0.5\textwidth}
  \includegraphics[width=.8\linewidth]{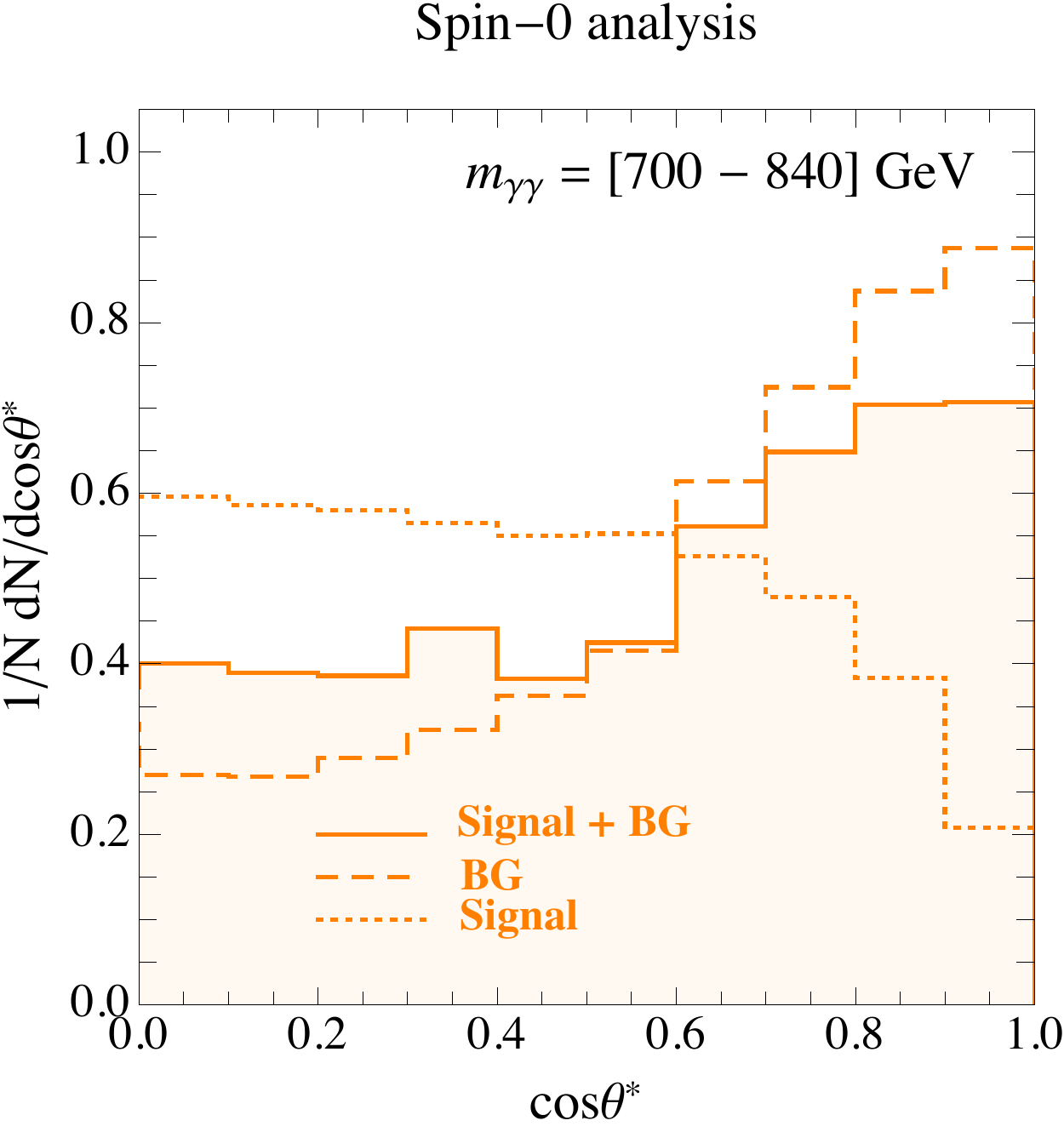}
\endminipage\hfill
\minipage{0.5\textwidth}
  \includegraphics[width=.8\linewidth]{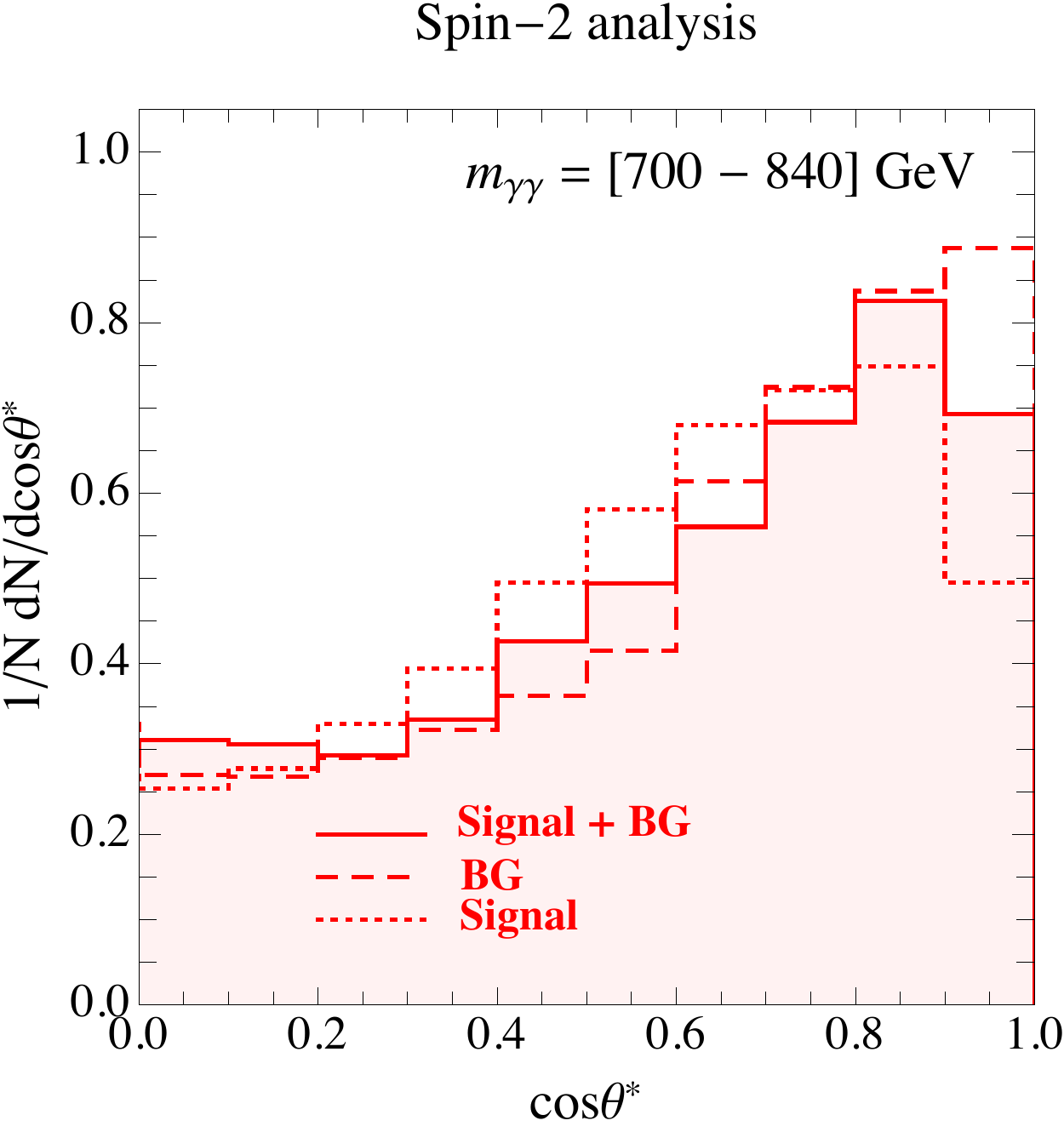}
\endminipage
\caption{\em 
Angular distributions (normalised to $1$ in the interval $\cos\theta^* \in [-1,1]$) for the loose set of selection cuts 
for the spin-0 (left panel, orange) and spin-2 (right panel, red) analysis.
}
\label{fig:Histograms2}
\end{figure}

In fig.~\ref{fig:Histograms} (fig.~\ref{fig:Histograms2}) we show 
the pdf's for the tight  (loose) selection cuts.
The left (right) panels refer to the spin-0 (spin-2) signal hypothesis.
We focus on the signal region with invariant mass $m_{\gamma\gamma} = [700-840]$ GeV, 
and we compare angular distributions for signal (dotted lines), background (dashed lines) and signal-plus-background (solid lines) simulated events.

As already discussed, the tight selection 
cuts 
have the effect of suppressing the forward region with $|\cos\theta^*| \gtrsim 0.7$. On the contrary, 
the loose selection cuts in eq.~(\ref{eq:LooseCuts}) populate the forward region. This is generally true for background and signals alike irrespective of the spin of the signal. In other words, an enhanced number of events in the forward direction is, by itself,  only the effect of the loosening of the selection cuts and provides no indication about the spin.

\subsection{Log likelihood ratio}\label{sec:LLRanalysis}

The PDFs derived in section~\ref{sec:PDFs} can be used to test 
the discriminative power of a LLR  between the spin-0 and 2 distributions.
For a given pdf describing the $\cos\theta^{*}$ distribution---either signal, background or a mixture of both---it is possible to randomly generate 
$N_{\rm obs}$ events and compute the LLR
\be
2 \log \frac{ \cal{L}_{\rm spin-0} }{ \cal{L}_{\rm spin-2} } \, . \label{eq:LLR}
\ee
 By repeating this {\it pseudo-experiment} $N_{ps}$ times, 
it is possible  to construct a  sample 
that can be used to compute the statistical distribution 
of a certain hypothesis. 

Let us  take $N_{\rm obs}^{(J)}$ spin-$J$  signal 
events
generated according to the corresponding pdf (discussed in section~\ref{sec:PDFs}) as well as $N_{\rm obs}^{(\rm bkg)}$ events. Each event $i$
is characterised by the value of the cosine of the CS scattering angle $z_i$ defined in eq.~(\ref{eq:CSdefinition}).
The likelihood function for the spin hypothesis $J^{\prime}$
is given by
\begin{equation}\label{eq:Likelihood}
\mathcal{L}_{{\rm spin}-J^{\prime}} = e^{-N_{\rm obs}^{(J)} - N_{\rm obs}^{(\rm bkg)}}
\prod_{i = 1}^{N_{\rm obs}^{(J)} + N_{\rm obs}^{(\rm bkg)}}
\left[
N_{\rm obs}^{(J)}\times {\rm pdf}_{J^{\prime}}(z_i)  + N_{\rm obs}^{(\rm bkg)}\times {\rm pdf}_{\rm bkg}(z_i) 
\right]~.
\end{equation}
As mentioned above, by repeating this measurement $N_{ps}$ time is possible to construct a statistical sample for $\mathcal{L}_{J^{\prime}}$.
We use $N_{ps} = 10^4$. 

We follow this prescription and  compute the LLR  in eq.~(\ref{eq:LLR}) using the definition in eq.~(\ref{eq:Likelihood}).
For each of the cases relevant for our analysis (see discussion below) 
we construct two statistical samples for the ratio $2 \log(\mathcal{L}_{\rm spin-0}/\mathcal{L}_{\rm spin-2})$: the first one populated with events
generated  according to the spin-0 distribution, the second one with spin-2 events.    
We expect the distribution of the ratio $2 \log(\mathcal{L}_{\rm spin-0}/\mathcal{L}_{\rm spin-2})$
to be peaked at positive values for events generated according to the spin-0 pdf, since in average $\mathcal{L}_{\rm spin-0} > \mathcal{L}_{\rm spin-2}$.
We expect a distribution peaked towards negative values for events generated according to the spin-2 pdf, 
since in this case $\mathcal{L}_{\rm spin-0} < \mathcal{L}_{\rm spin-2}$.

We explore different cases:
\begin{itemize}

\item[$\circ$]  $N_{\rm obs}^{(\rm bkg)} = 0$

 In this case we consider only signal events.
 This ideal situation applies to the experimental data after background subtraction. 
 We already stressed in section~\ref{sec:sPlot} that this is a complicated task, and we 
 proposed the $_s\mathcal{P}lot$ technique to tackle the issue.
 The result obtained in this case should be considered as
  the case in which the signal has been separated from the background without any loss of information. It is an ideal case that gives the best discrimination one can possibly achieve.

\item[$\circ$]  $N_{\rm obs}^{(J)} = N_{\rm obs}^{(\rm bkg)}$

In this case we add to the signal events an equal number of background events.
This case captures  the impact of the systematic error due to the contamination of the background in the signal sample, as well as the uncertainties in the background modelling.

\end{itemize}
In each of these two cases we consider the analysis with both 
tight and loose selection cuts.

\begin{figure}[!htb!]
\begin{center}
\fbox{Log likelihood ratio (signal only)}
\end{center}
\centering
\minipage{0.5\textwidth}
  \includegraphics[width=.8\linewidth]{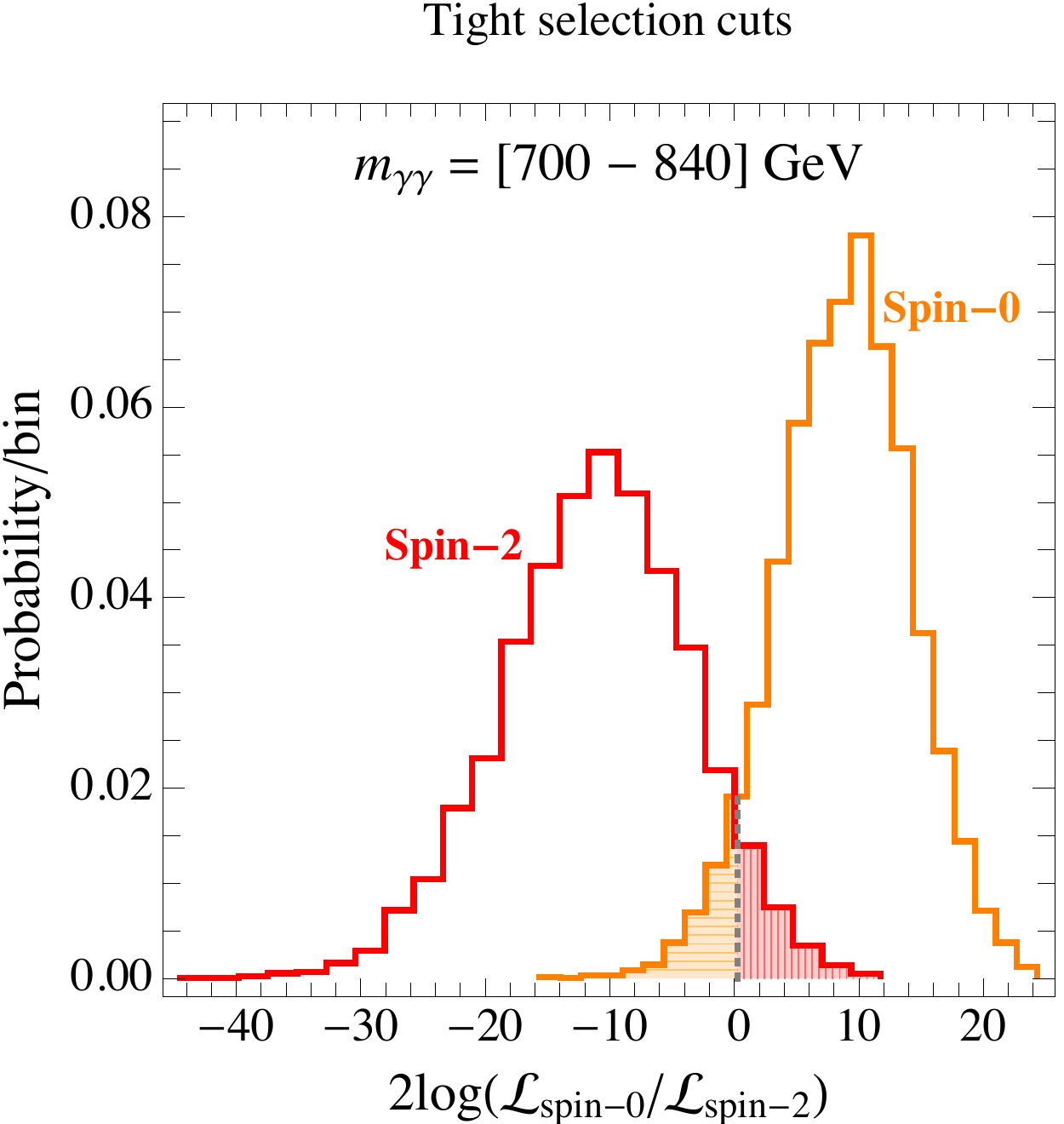}
\endminipage\hfill
\minipage{0.5\textwidth}
  \includegraphics[width=.8\linewidth]{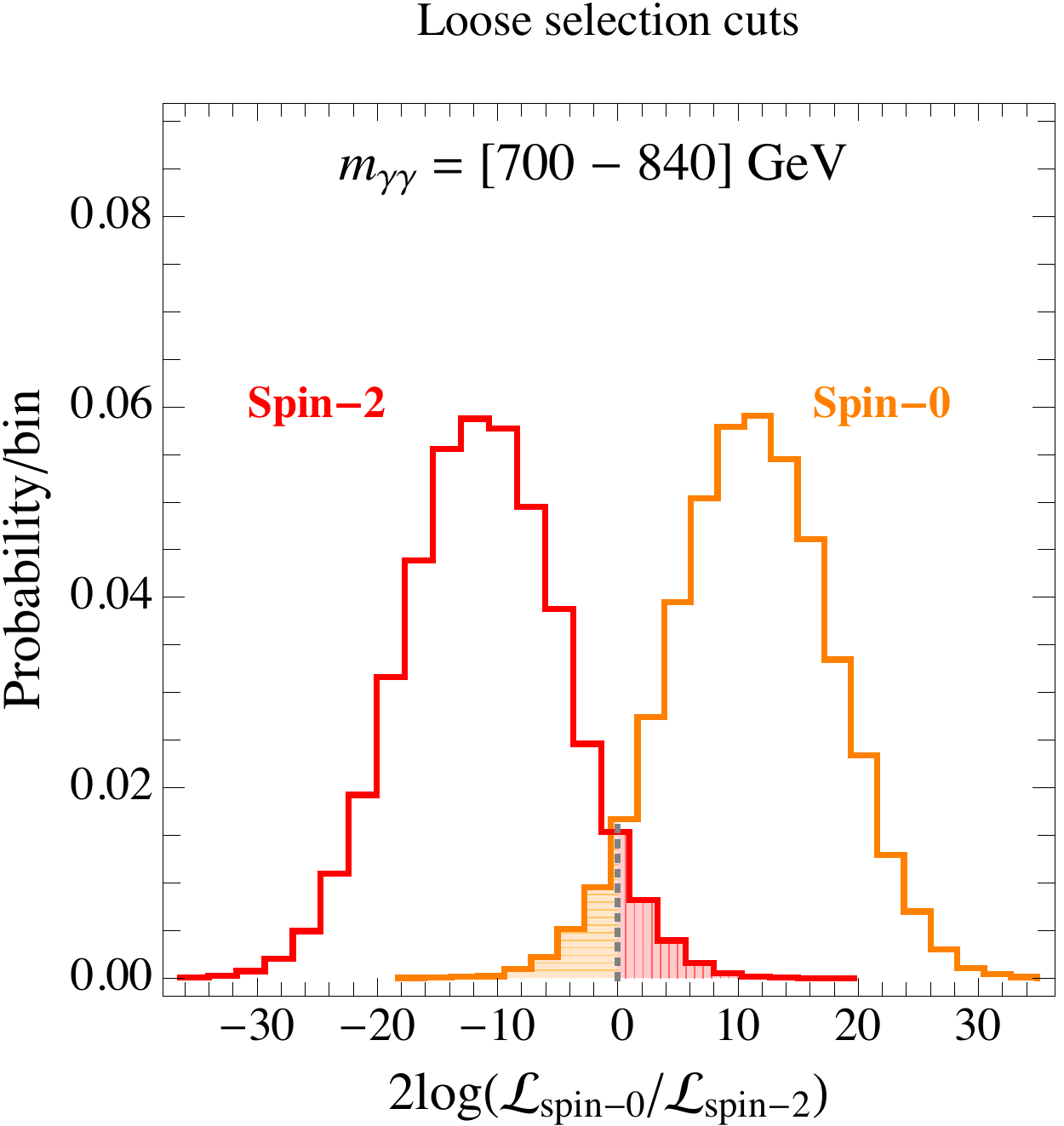}
\endminipage
\caption{\em 
LLR for $N_{\rm obs}^{(J)} = 40$ events and $N_{ps} = 10^4$ simulated pseudo-experiments. 
We show in the left panel (right panel) the analysis with tight selections cuts (loose selection cuts).
We included in our Monte Carlo simulation only the distributions for the signal. 
The 
vertical grey line marks the point at which the two areas in red (with vertical meshes, for the spin-2 distribution) and orange (with horizontal meshes, for the spin-0 distribution) are equal.  
}
\label{fig:LikSignal}
\end{figure}

\begin{figure}[!htb!]
\begin{center}
\fbox{Log likelihood ratio (signal = background)}
\end{center}
\centering
\minipage{0.5\textwidth}
  \includegraphics[width=.8\linewidth]{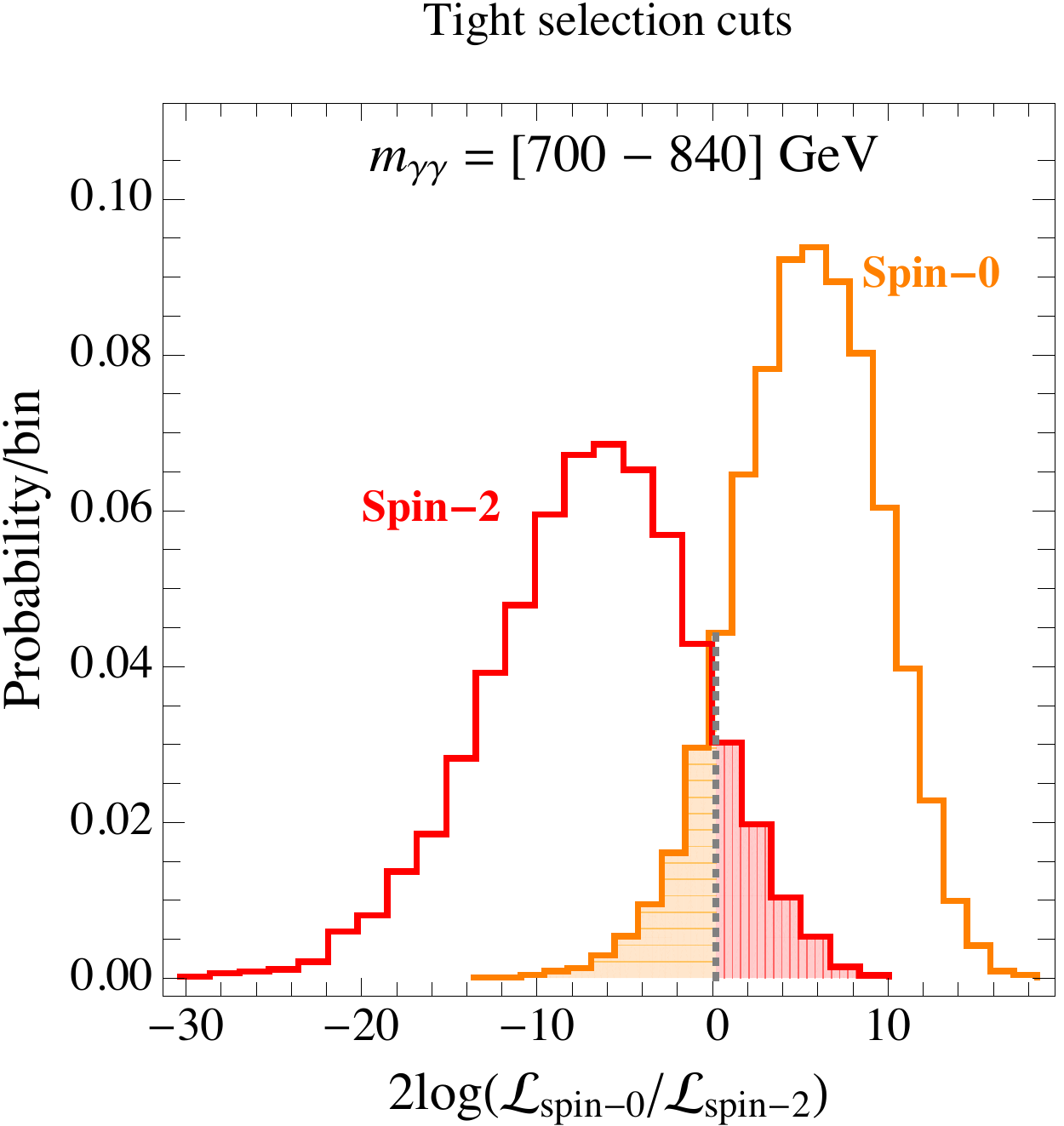}
\endminipage\hfill
\minipage{0.5\textwidth}
  \includegraphics[width=.8\linewidth]{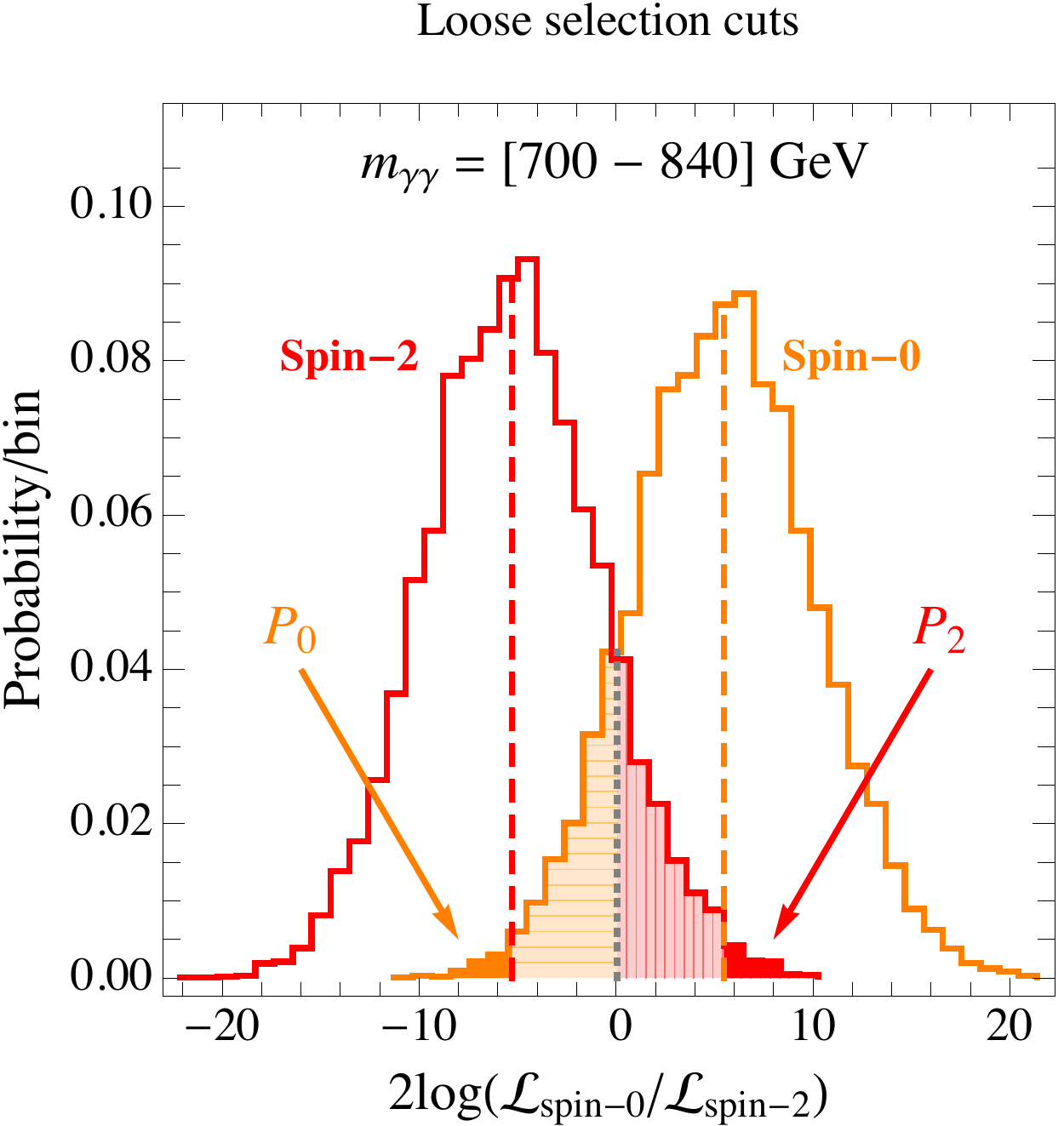}
\endminipage
\caption{\em 
The same as in fig.~\ref{fig:LikSignal} but with
$N_{\rm obs}^{(J)} = N_{\rm obs}^{(\rm bkg)} = 40$ events for both background and signal. 
}
\label{fig:LikSignalBG}
\end{figure}

In fig.~\ref{fig:LikSignal}  (fig.~\ref{fig:LikSignalBG}) we show our results for the LLR analysis 
considering the case with only signal events (with an equal number of signal and background events).
The left (right) panel refers to tight (loose) section cuts.
We take $N_{\rm obs}^{(J)} = 40$ ($N_{\rm obs}^{(J)} = N_{\rm obs}^{(\rm bkg)} = 40$) for the analysis with only signal events (with an equal number of signal and background events).

The LLR in figs.~\ref{fig:LikSignal},\,\ref{fig:LikSignalBG}
take the form of two distinct bell-shaped distributions
when computed  for the spin-0 and spin-2 hypotheses.
To quantify the difference in terms of statistical significance 
we  first compute the point beyond which the right-side tail of the left 
distribution and the left-side tail of the right one have equal areas. This point corresponds to a hypothesis test in which no preference for the spin of the resonance is assumed. 
We mark the point with a vertical dotted grey line. 
The  two equal areas correspond to a $p$-value which can be translated into a significance  $\mathcal{Z}$ as
\be
\mathcal{Z}= \Phi^{-1}(1-p) \, , \label{erf}
\ee
where 
\be
\Phi(x) = \frac{1}{2} \left[ 1 + \erf \left( \frac{x}{\sqrt{2}} \right)\right] \, .
\ee
We  use the value of $\mathcal{Z}$ to 
assign a statistical significance to the separation between the two LLR distributions. The significance can be turned into $\mathcal{Z} \sigma$, the number of $\sigma$'s,  in the approximation in which the distribution is assumed to be Gaussian.

When an actual experiment is performed, a particular value of LLR is obtained. The associated $p$-value can be computed and the significance of the observation estimated according to the same procedure. For instance, in the right side of fig.~\ref{fig:LikSignalBG}, the $p$-value $P_0$ ($P_2$) would correspond to a test of the spin-0 (spin-2) hypothesis  after the experiment has produced a value close to the corresponding medians. In the approximation of Gaussian distributions, the significance of the given hypothesis corresponds to twice the $\mathcal{Z}$ of the separation.


In fig.~\ref{fig:LikSignal} the analysis with only signal events allows for a clear separation between the two hypotheses, with 
 the loose selection cuts (right panel)
performing slightly better. This is to be expected, since the loose cuts populate the forward region, thus 
increasing the discriminating power between the two cases.
If quantified in terms of the significance $\mathcal{Z}$ 
introduced before, we find $\mathcal{Z} = 1.56$ ($\mathcal{Z} = 1.67$) for tight (loose) selection cuts.
The inclusion of background events reduces the statistical significance 
of the analysis, and  we find $\mathcal{Z} = 1.23$ ($\mathcal{Z} = 1.18$) for tight (loose) selection cuts.

\subsection{Center-edge asymmetry} 

We define the center-edge asymmetry~\cite{Dvergsnes:2004tw}
\begin{equation}\label{eq:ACE}
\mathcal{A}_{CE} \equiv  
\frac{N_{C} - N_{E}}{N_{C} + N_{E}}~,
\end{equation}
where $N_{C}$ ($N_{E}$) is the number of events with $-z^{*} \leqslant \cos\theta^{*}  \leqslant  z^{*}$ ($|\cos\theta^{*}| > z^{*}$).
The parameter $z^*$ is a threshold parameter that can be tuned to optimise the separation between the
spin-0 and spin-2 cases.

The analysis for the asymmetry proceed in close analogy 
with what discussed in the case of the LLR.
As before, using the pdf's in section~\ref{sec:PDFs}  we 
construct two statistical samples for the center-edge asymmetry 
under the spin-0 and spin-2 hypothesis. 
Case by case, we can assign a statistical significance 
using the separation $\mathcal{Z}$ defined in section~\ref{sec:LLRanalysis}.

\begin{figure}[!ht]
\begin{center}
\fbox{Center-edge (signal only)}
\end{center}
\centering
\minipage{0.5\textwidth}
  \includegraphics[width=.8\linewidth]{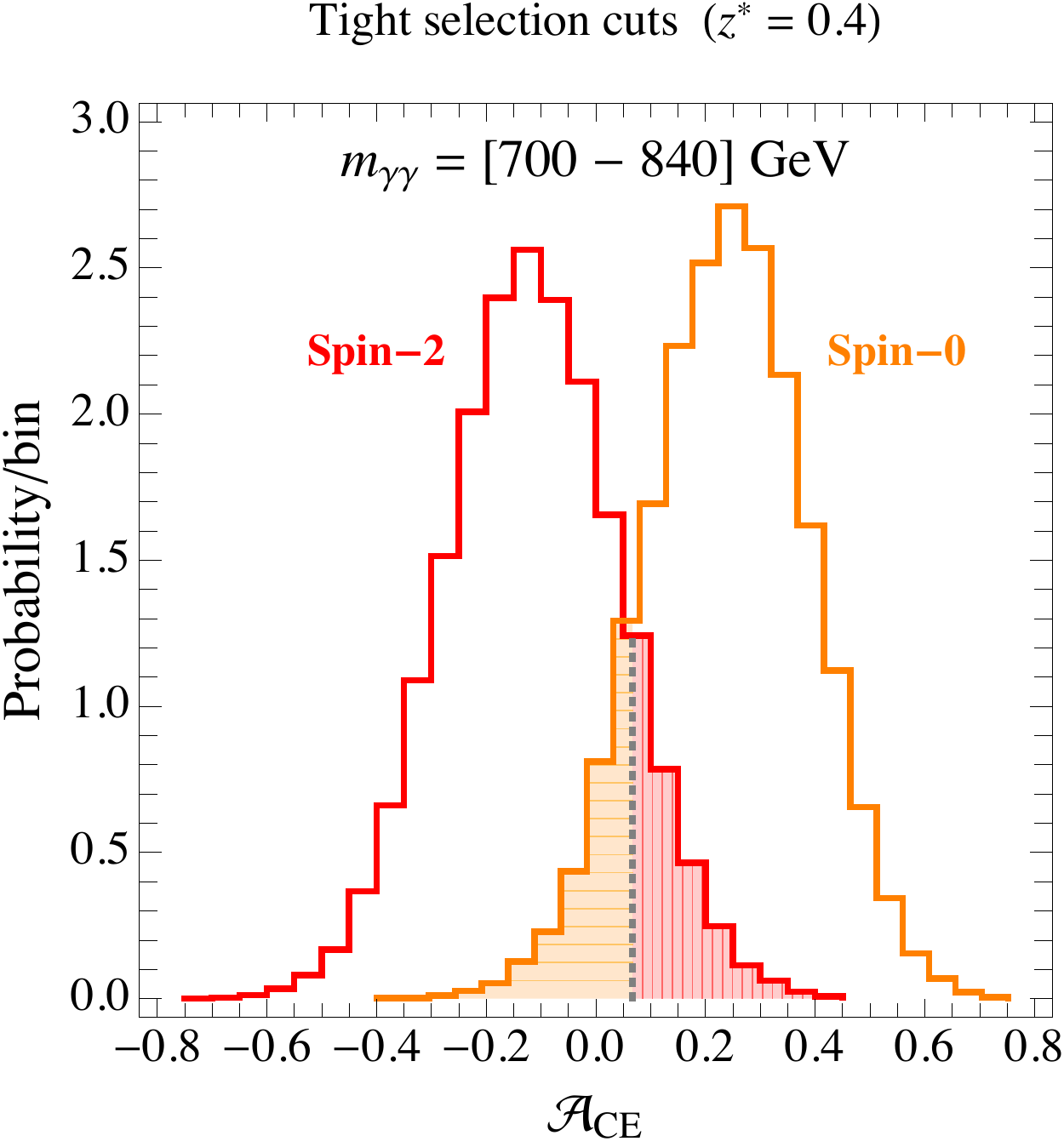}
\endminipage\hfill
\minipage{0.5\textwidth}
  \includegraphics[width=.8\linewidth]{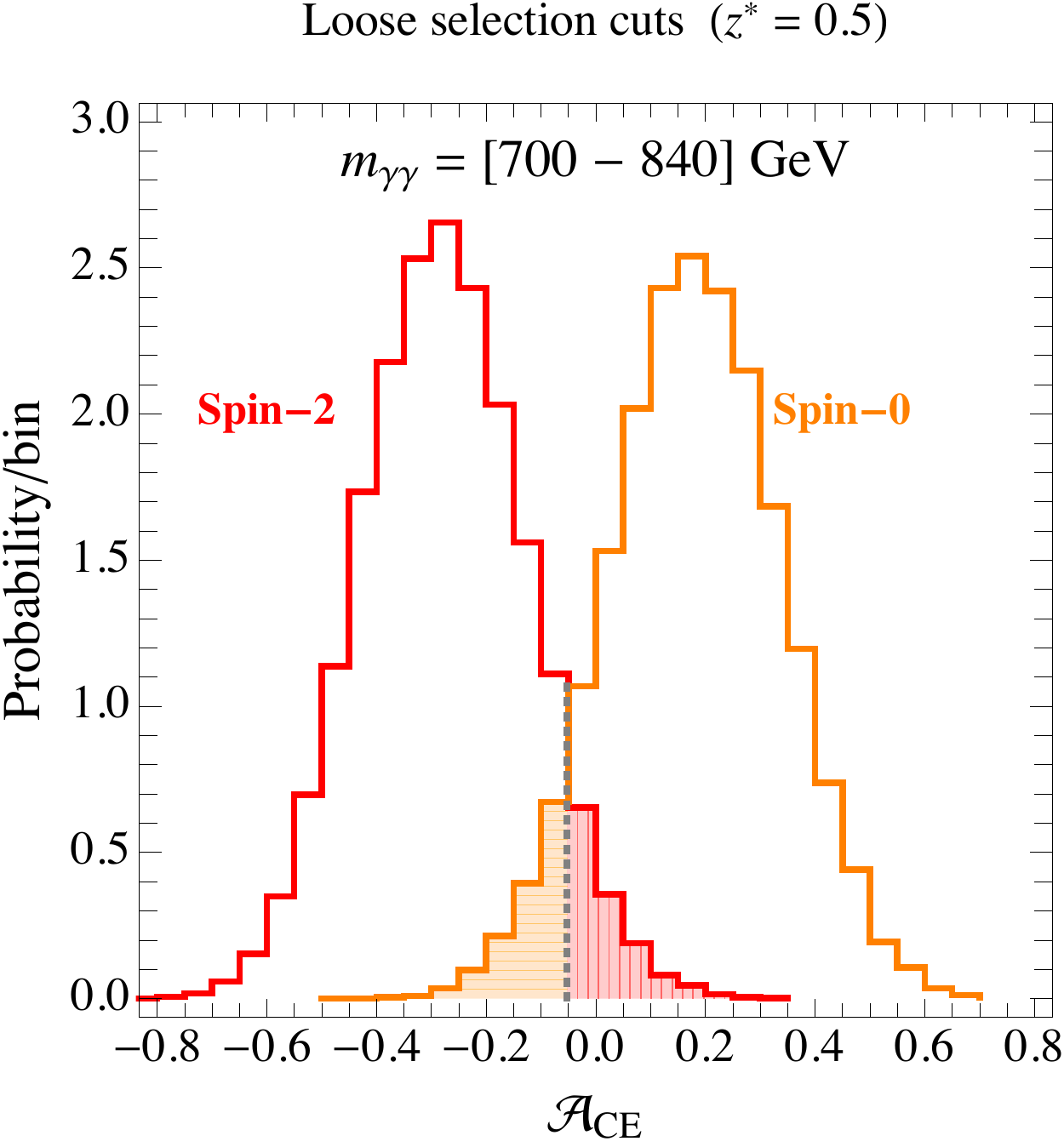}
\endminipage
\caption{\em 
Central-edge asymmetry distributions for $N_{\rm obs}^{(J)} = 40$ events and $N_{ps} = 10^4$ simulated pseudo-experiments. 
We show in the left panel (right panel) the analysis with tight selections cuts (loose selection cuts).
We included in our Monte Carlo simulation only the distributions for the signal. 
The 
vertical grey line marks the point at which the two areas in red (with vertical meshes, for the spin-2 distribution) and orange (with horizontal meshes, for the spin-0 distribution) are equal.  
}
\label{fig:ACESignal}
\end{figure}

\begin{figure}[!h]
\begin{center}
\fbox{Center-edge (signal = background)}
\end{center}
\centering
\minipage{0.5\textwidth}
  \includegraphics[width=.8\linewidth]{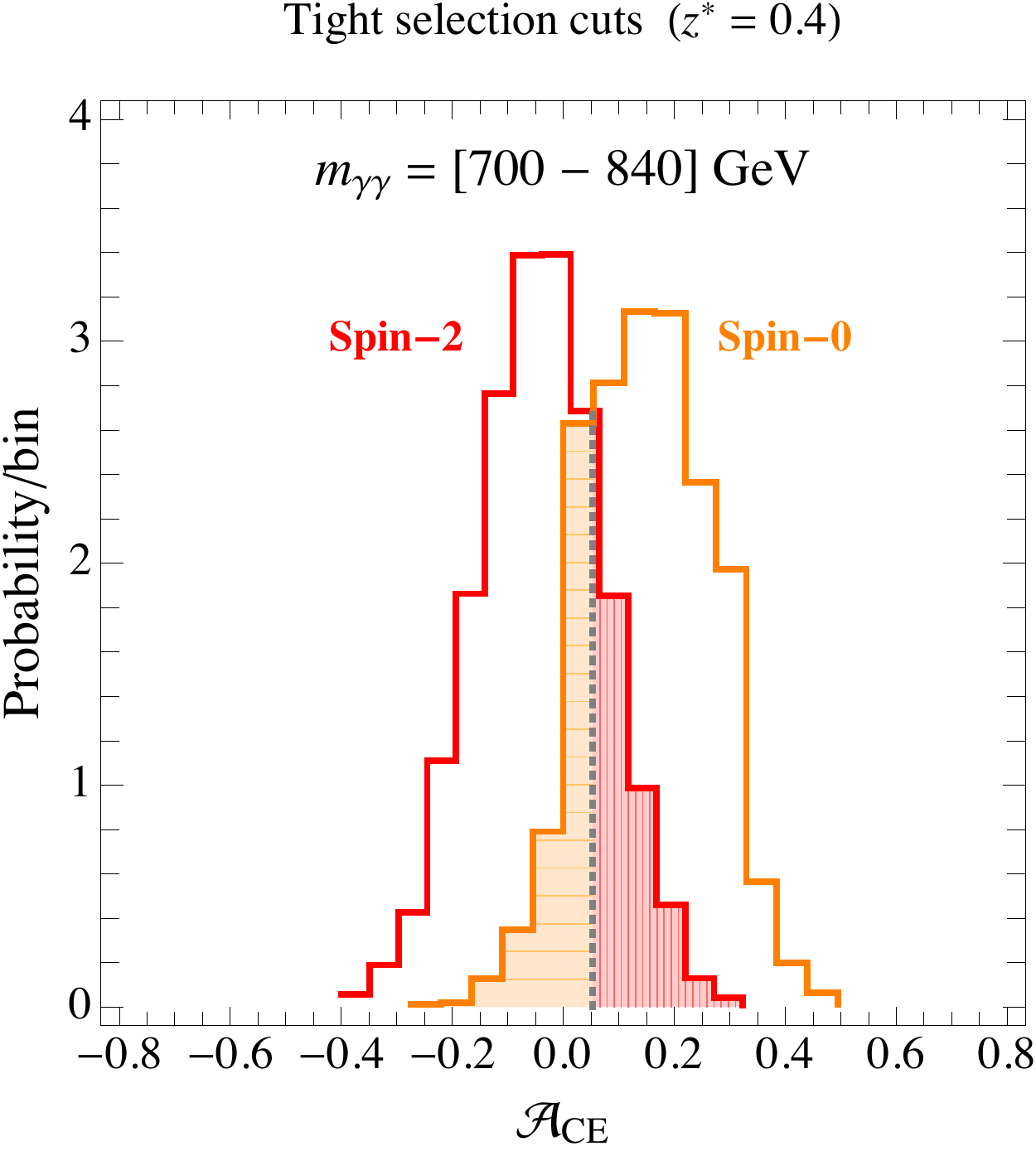}
\endminipage\hfill
\minipage{0.5\textwidth}
  \includegraphics[width=.8\linewidth]{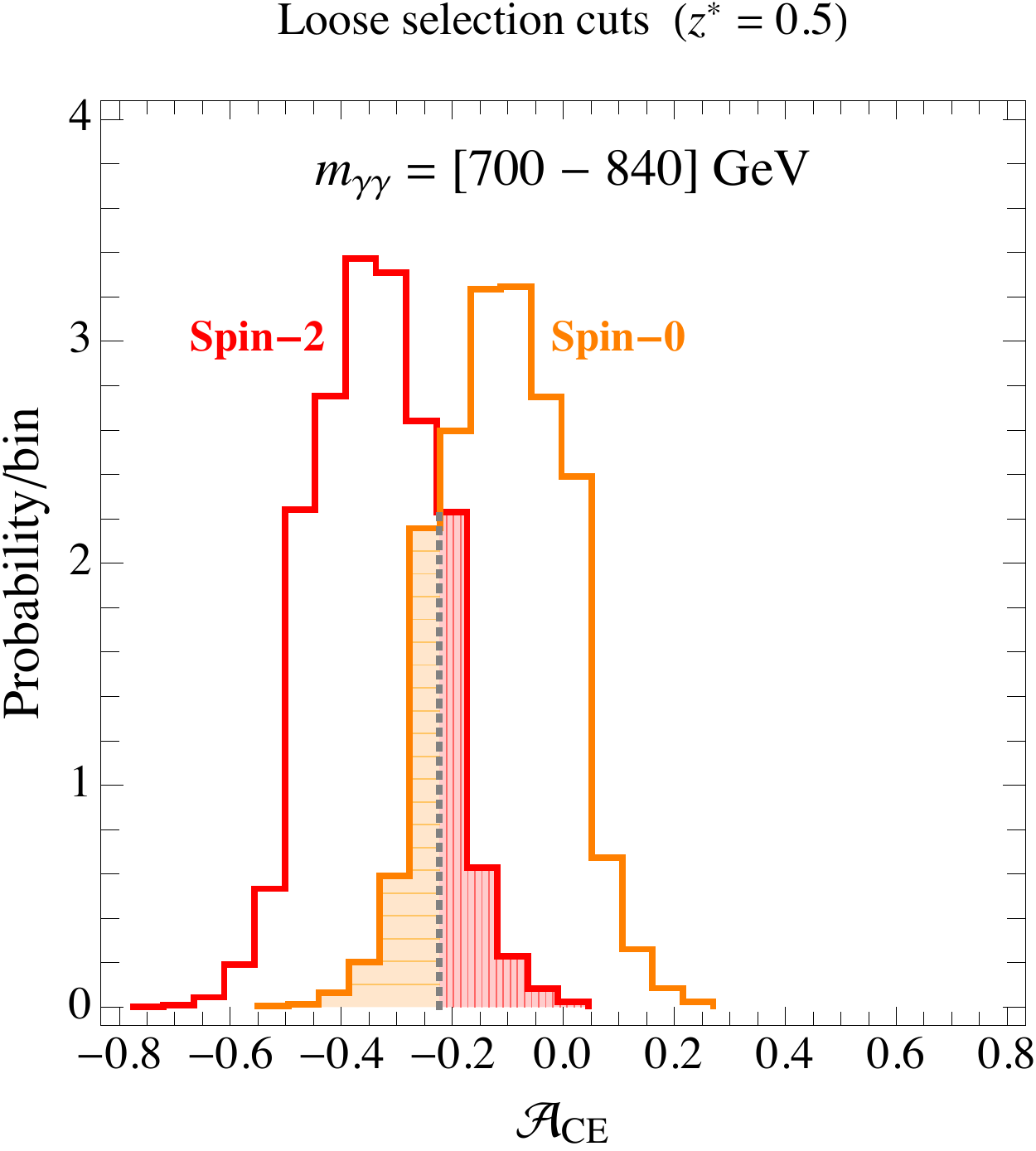}
\endminipage
\caption{\em 
The same as in fig.~\ref{fig:ACESignal} but with
$N_{\rm obs}^{(J)} = N_{\rm obs}^{(\rm bkg)}  = 40$ events for both background and signal. 
}
\label{fig:SigBGACESignal}
\end{figure}

We present our results in fig.~\ref{fig:ACESignal} (fig.~\ref{fig:SigBGACESignal}) for the case with $N_{\rm obs}^{(J)} = 40$ signal events (with an equal number of 
 signal and background events, $N_{\rm obs}^{(J)} = N_{\rm obs}^{(\rm bkg)}  = 40$). 
 As before, the left (right) panel refers to the analysis with tight (loose) cuts.
Furthermore, we find that the value $z^{\star} = 0.4$ is best suited for tight cuts while $z^{\star} = 0.5$ allows for a better separation if loose cuts.
are imposed.

The medians of the probability distributions in fig.~\ref{fig:SigBGACESignal} (and the similar ones below) do not correspond to their values at the partonic level---as computed by means of \eq{eq:AngoloSpin0} and \eqs{eq:Spin2gg}{eq:Spin2qq}---because of the distortions introduced by the detector.

Considering only signal events, we find the significance 
$\mathcal{Z} = 1.20$ ($\mathcal{Z} = 1.41$) for tight (loose) selection cuts.
 Loose selection cuts allow for a better separation between the two spin hypotheses.
We notice that the increase in significance gained going from tight to loose cuts 
is larger for the asymmetry if compared with what found for the  LLR.
This is because the asymmetry is, by definition, more sensitive to
the way in which the events are arranged as a function of the CS angle.

The significance $\mathcal{Z}$ decreases as a consequence of the inclusion of background events, and 
 we find $\mathcal{Z} = 0.80$ ($\mathcal{Z} = 0.98$) for tight (loose) selection cuts.
Notice that for loose selection cuts 
the inclusion of background events 
shifts the central value of the spin-0 distribution towards negative values.
This is because the background, dominated by photon pairs produced in $q\bar{q}$ annihilation, 
is peaked  in the forward region as a consequence of $t-$ and $u$-channel quark exchange (see figs.~\ref{fig:Histograms},\,\ref{fig:Histograms2}, right panels).
As a result, a large background  contamination of the spin-0 signal sample distorts the distribution 
towards the forward region, and eventually  leads to negative values for the central-edge asymmetry---especially if loose cuts are imposed.


\subsection{Significance vs.\ number of events}
\begin{figure}[!h]
\centering
\minipage{0.5\textwidth}
  \includegraphics[width=.8\linewidth]{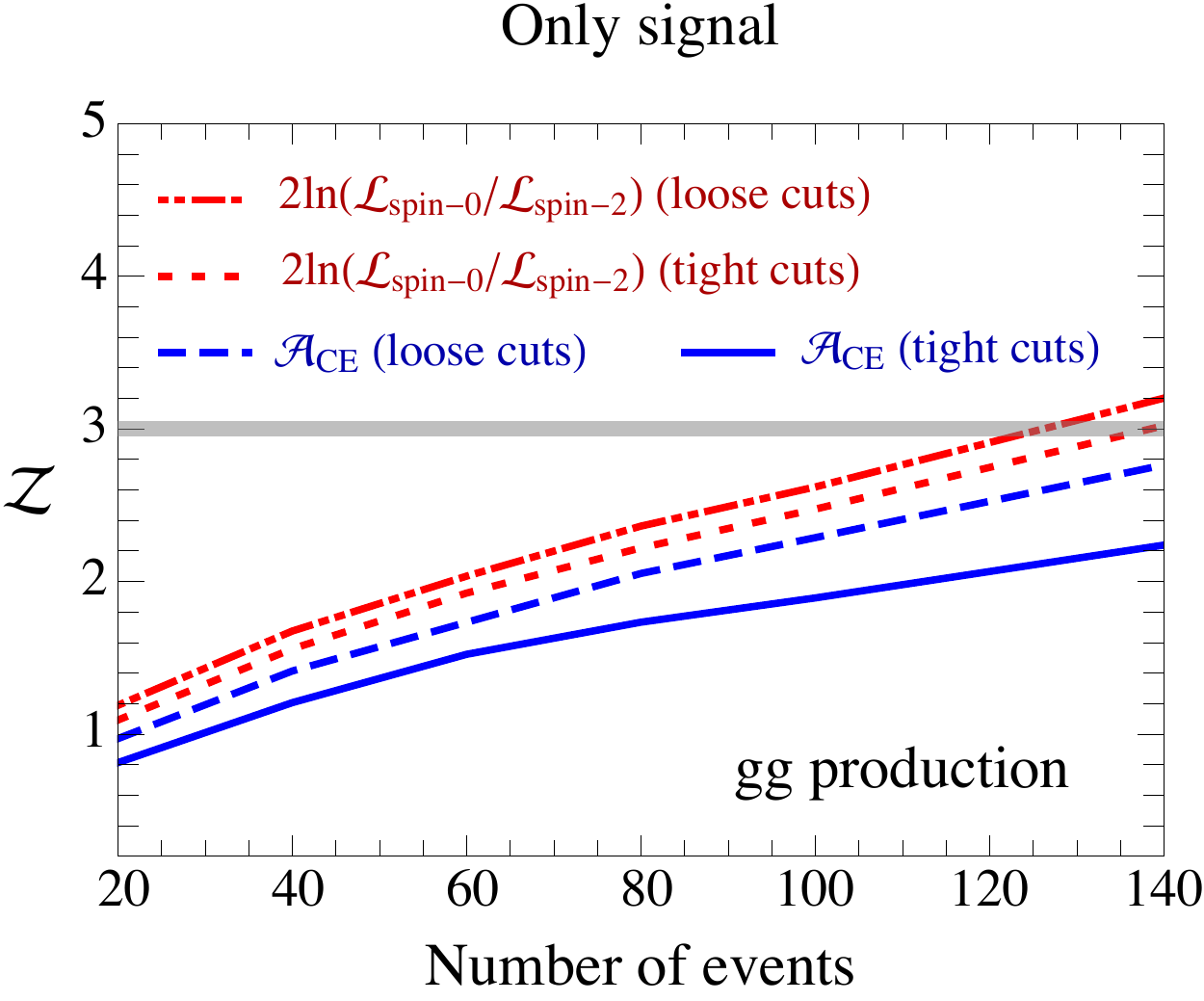}
\endminipage\hfill
\minipage{0.5\textwidth}
  \includegraphics[width=.8\linewidth]{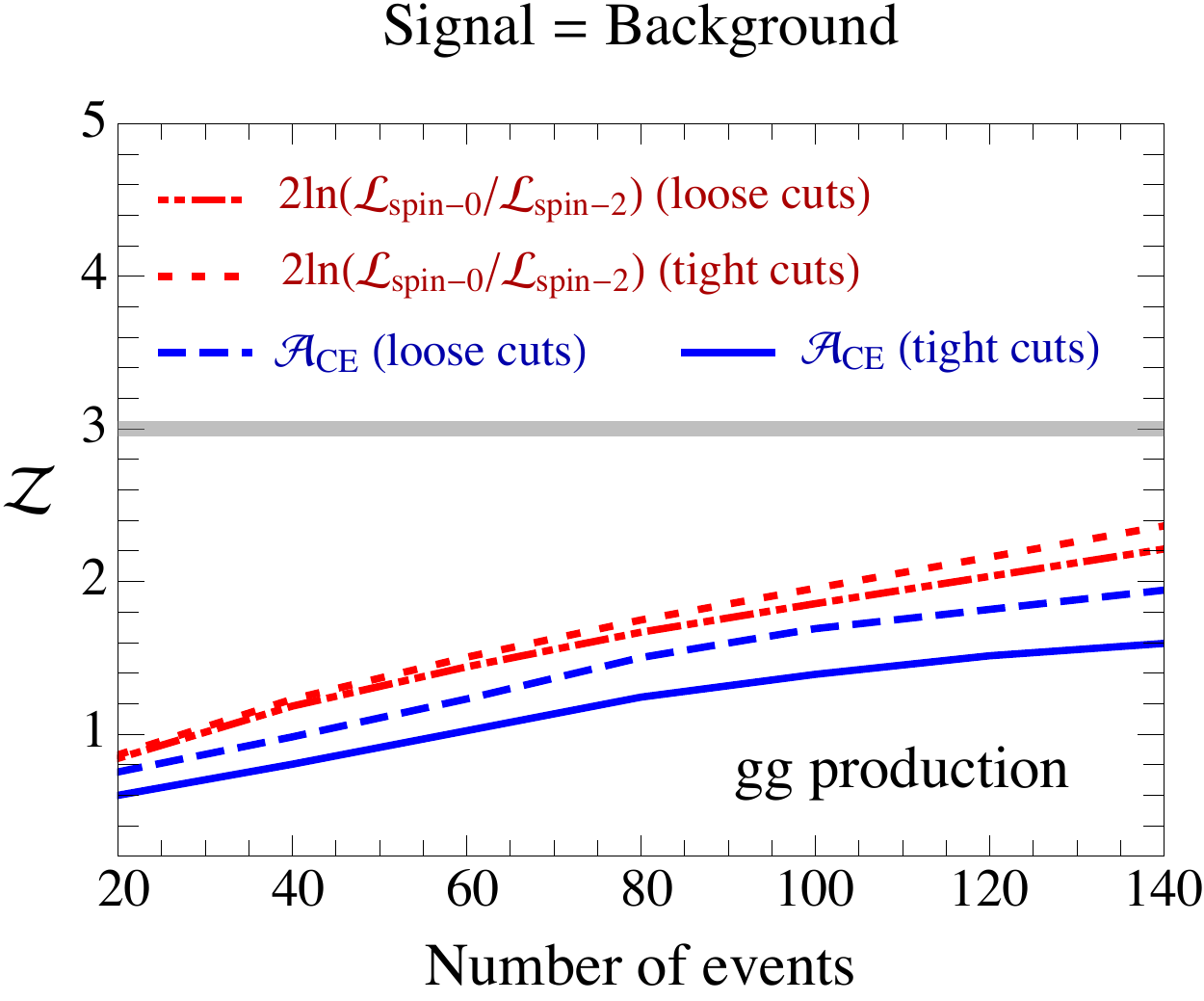}
\endminipage
\caption{\em Significance $\mathcal{Z}$ for the LLR and the center-edge asymmetry as a function of the number of events. In the model for the spin-2 resonance, the production mechanism is taken to be gluon fusion only.
In this figure the significance does not refer to one particular spin hypothesis but to the separation between the two distributions (eq.~(\ref{erf}) 
computed with the p-value representing the area equally shared between the two LLRs---see caption of fig.~\ref{fig:ACESignal}).
The significance of one particular spin hypothesis---eq.~(\ref{erf}) computed with the p-value representing one of the two regions, $P_0$ or $P_2$, in fig.~\ref{fig:LikSignalBG}---is larger by a factor of 2. This case is discussed in section~\ref{sec:KKGraviton} when the  spin-2 hypothesis of the KK graviton will be tested.
}
\label{fig:MoneyPlot}
\end{figure}

All the results presented and discussed so far 
were based on a fixed value of observed events.
From a more general perspective,  the most valuable  information  that can be extracted from the analysis is
 the number of events necessary to reach a certain level of significance $\mathcal{Z}$ in separating the different spin hypotheses
 This number is shown in fig.~\ref{fig:MoneyPlot} for the two cases, signal only and signal plus background, under discussion. From fig.~\ref{fig:MoneyPlot} we see that starting with 60 events and a significance around 2 for the case of LLR on the signal only,  we can reach a significance around 3 by simply doubling the number of events.  A similar improvement in significance can be seen in the other, less optimal, cases.
 
 Even though a comparison of the two approaches shows (as it should be expected~\cite{NP}) that the LLR performs roughly 20\% better than the central-edge asymmetry, one might bear in mind that the asymmetry is more robust against possible unknown uncertainties and provides a clearer physical picture.
 
 Let us stress that the significance in fig.~\ref{fig:MoneyPlot} refers to the separation between the two distributions, and it was computed by means of eq.~(\ref{erf}) 
with  the p-value representing the area equally shared between the two LLRs (defined by the gray dashed vertical line in figs.~\ref{fig:LikSignal}-\ref{fig:SigBGACESignal}).
The plot in fig.~\ref{fig:MoneyPlot} can be used to directly test  the two spin hypotheses by multiplying by a factor of 2 the significance.  Assuming an efficient background subtraction and a production of the spin-2 dominated by gluon fusion, roughly 40 events are necessary to reach a significance  $\mathcal{Z}=3$ by means of the LLR. This value is mostly independent of the selection cuts used.  Given the cross section value obtained by the fit on the invariant mass distribution in section \ref{sec:Fit}---this number of events corresponds to a luminosity of around 5 (7) fb$^{-1}$ after loose (tight) selection cuts.  Therefore a luminosity of 10 fb$^{-1}$ will make it possible to determine the spin of the new resonance. This is also a luminosity for which the $5\, \sigma$ significance threshold for a discovery has been comfortably crossed.

\section{The case of the lightest Kaluza-Klein graviton excitation}
 \label{sec:KKGraviton}

Scenarios with extra dimensions generically predict the existence of massive spin-2 particles ($G^*$ hereafter), corresponding to the KK excitations of the graviton.
 A possible signature of these models (in particular in the presence of strongly-warped extra dimensions at the TeV scale) 
 could be the discovery of a single resonance---the lightest excitation in the KK tower $G^*$ mentioned above---whose phenomenology can be effectively  
 described by its mass and its universal coupling with the SM energy-momentum tensor (see eq.~(\ref{eq:LagSpin2})).
 
\begin{figure}[!ht]
\centering
  \includegraphics[width=.5\linewidth]{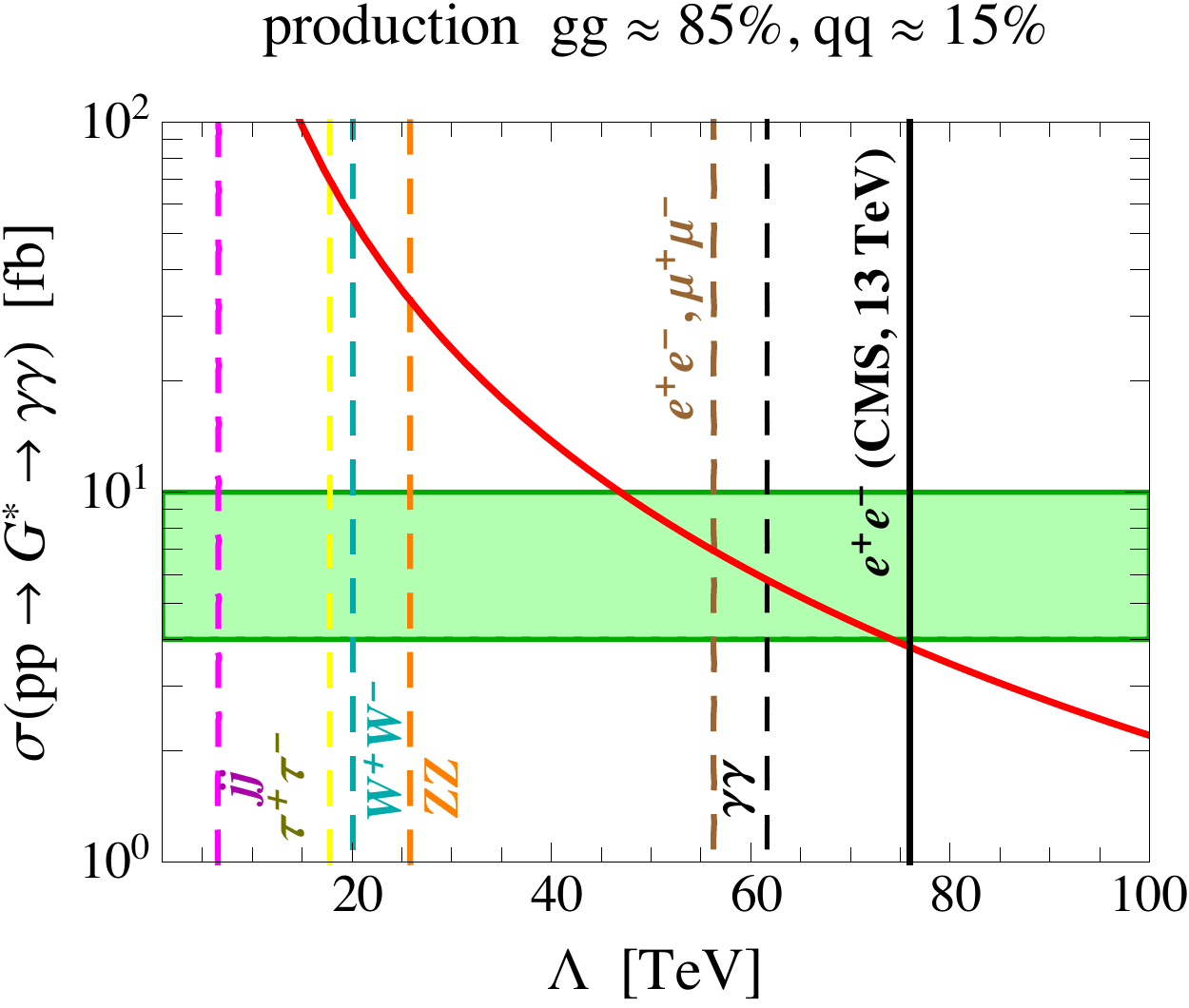}
\caption{\em 
Parameter space for the lightest KK spin-2 graviton excitation $G^*$.
We show in red the di-photon cross-section $\sigma(pp\to G^* \to \gamma\gamma)$ at $\sqrt{s} = 13$ TeV as a function
of the effective scale $\Lambda$.
The vertical dashed lines mark the regions excluded at $\sqrt{s} = 8$ TeV.
The vertical solid line corresponds to the CMS bound from $e^+ e^-$ final state at $\sqrt{s} = 13$ TeV~\cite{Giddings:2016sfr}.
The green band reproduces the cross-section $\sigma(pp\to G^* \to \gamma\gamma) = [4 -10]$ fb.
}
\label{fig:KKParameterSpace}
\end{figure}

 The validity of this setup in the light of the 750 GeV excess was discussed in~\cite{Giddings:2016sfr} (see also~\cite{Carmona:2016jhr} for related works).
 The situation is summarised in fig.~\ref{fig:KKParameterSpace} where we show the allowed parameter space (see caption for details).
For $M_{G^*} = 750$ GeV 
the representative di-photon cross-section $\sigma(pp\to G^* \to \gamma\gamma) = [4-10]$ fb at $\sqrt{s} = 13$ TeV---necessary to fit the observed excess---implies 
$\Lambda \approx [47-74]$ TeV. The corresponding total decay width is $\Gamma_{G^*} \approx [7-18]$ MeV.
The strongest bound on the model comes from di-lepton data at $\sqrt{s} = 13$ TeV~\cite{CMS:2015nhc}.
These bounds refer to searches for di-lepton decay of a spin-1 resonance
but they can be recast for the case of the KK graviton as shown in~\cite{Giddings:2016sfr}.
The $e^+e^-$ di-lepton final state in CMS 
put the constraint $\Lambda > 76$ TeV. The latter shows some tension with the signal strength needed to fit the di-photon excess at $\sqrt{s} = 13$ TeV.

Given the current uncertainties in the actual value of the signal strength needed to reproduced the observed excess, no 
strong conclusions can be derived from these bounds.~\footnote{In~\cite{750} the CMS collaboration presented a combined fit 
of di-photon data at at $\sqrt{s} = 8$ and $\sqrt{s} = 13$ TeV, and 
the reported best-fit value is  
$\sigma(pp\to G^* \to \gamma\gamma) = 4.5^{+1.9}_{-1.7}$ fb ($\Lambda = 86_{-17}^{+20}$ TeV)
that is compatible with the di-lepton bound.
Notice that our analysis does not depend on the specific value of $\Lambda$ assumed, since it cancels out 
from the normalised pdf of the signal.
} Nevertheless, it goes without saying that if the di-photon excess will be confirmed in the next future, 
the explanation in terms of the lightest KK graviton excitation can be ruled out in the absence of a corresponding bump in the di-lepton spectrum.
Keeping this discussion in mind, 
in the rest of this section we explore the KK graviton an example 
for our spin-2 analysis.

As far as the production mechanism is concerned, 
we find that gluon fusion accounts for about $85\%$ of the total rate while
 $q\bar{q}$  annihilation is responsible for the remaining  $15\%$.
 This is an important point, since it means that the angular distribution in eq.~(\ref{eq:Spin2gg})
 is  contaminated by the one in eq.~(\ref{eq:Spin2qq}). 
 The results presented in section~\ref{sec:Results}---strictly valid only in the case of production via gluon fusion---are therefore non longer applicable to a KK graviton, and a dedicated analysis is needed.
 

\begin{figure}[!htb!]
\centering
\minipage{0.5\textwidth}
  \includegraphics[width=.8\linewidth]{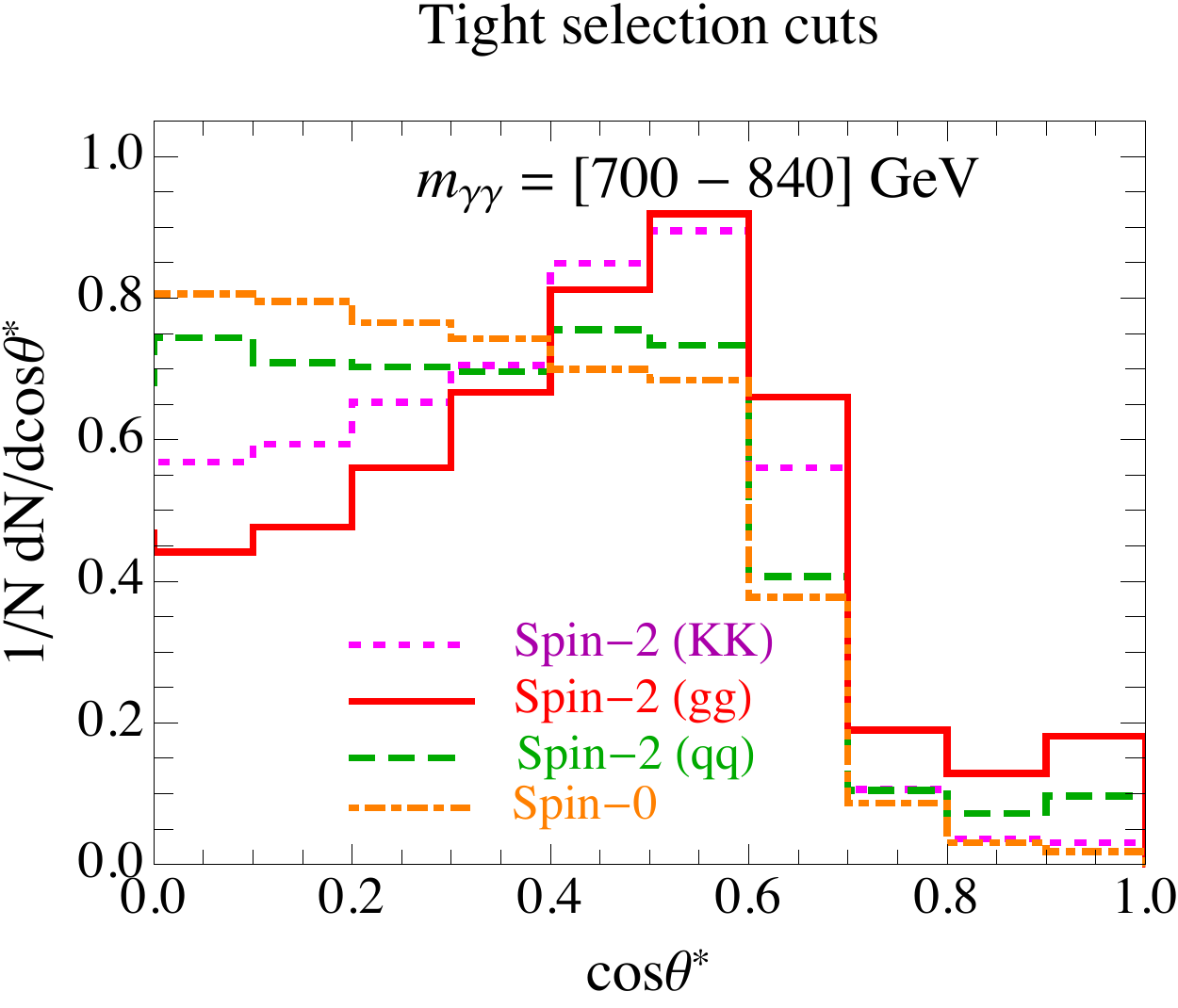}
\endminipage\hfill
\minipage{0.5\textwidth}
  \includegraphics[width=.8\linewidth]{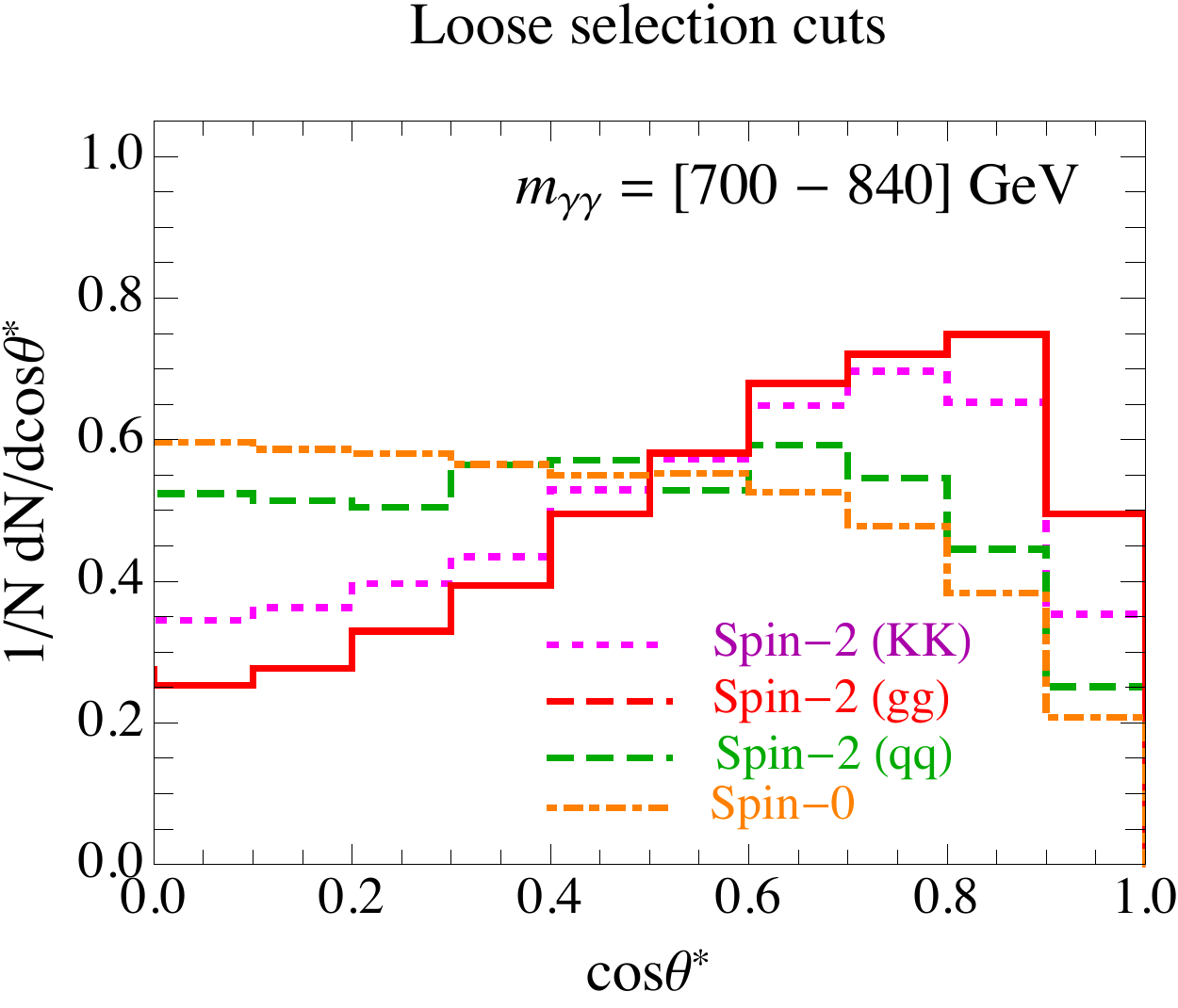}
\endminipage
\caption{\em 
Comparison between the angular distribution 
for the signal events. Dot-dashed orange: spin-0 resonance; 
solid red: spin-2 resonance produced via gluon fusion;
dashed green: spin-2 resonance mostly produced via $q\bar{q}$;
dotted magenta: spin-2 KK graviton with universal couplings.
We show the case with tight selection cuts in the left panel, and the analysis with loose selection cuts in the right one.
}
\label{fig:SpinqQuark}
\end{figure}

Following the same framework of section~\ref{sec:Methods} and \ref{sec:Results}, we show in fig.~\ref{fig:SpinqQuark}  the angular distributions 
for the signal samples corresponding to tight cuts (left panel) and loose cuts (right panel).
In addition to the spin-2 resonance produced via gluon fusion (solid red) and the spin-0 case (dot-dashed orange) discussed in section~\ref{sec:Results},
we show in dotted magenta the angular distribution for the KK graviton and in dashed green the case of a spin-2 resonance mostly produced 
by $q\bar{q}$ annihilation (we shall discuss this case in more detail in the next section).
These angular distributions reflect
what already expected from our general discussion.
The $15\%$ $q\bar{q}$ contamination 
for a KK graviton sizeably alters the pdf generated by gluon fusion.
As evident from the comparison with the dashed green line, the larger the $q\bar{q}$ contamination the closer the result to the spin-0 case.
It is now important to quantify this effect in terms of significance.

\begin{figure}[!htb!]
\centering
\begin{center}
\fbox{Center-edge (signal only)}
\end{center}
\minipage{0.5\textwidth}
  \includegraphics[width=.8\linewidth]{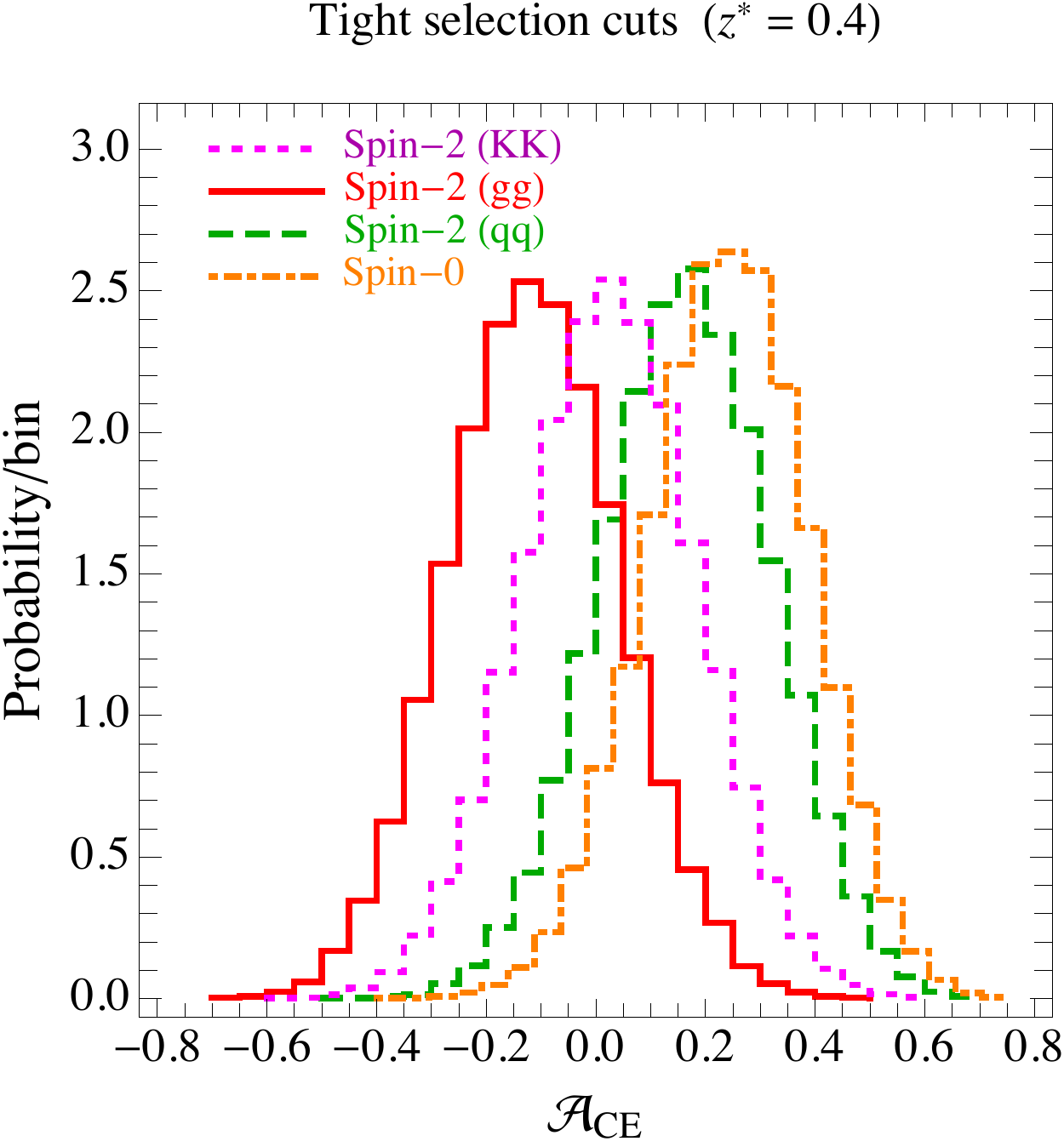}
\endminipage\hfill
\minipage{0.5\textwidth}
  \includegraphics[width=.8\linewidth]{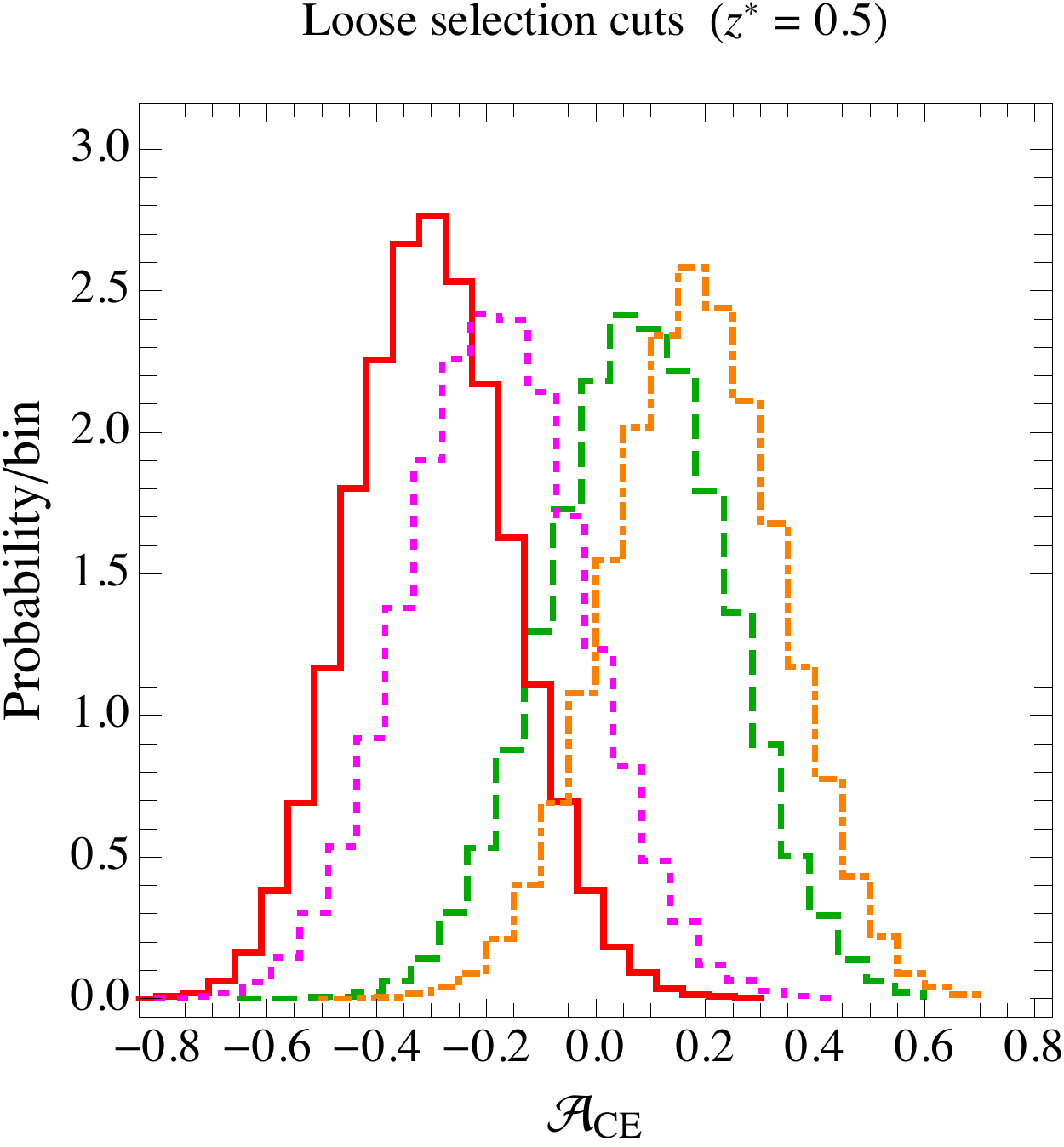}
\endminipage
\caption{\em 
The same as in fig.~\ref{fig:SigBGACESignal} but including also 
the central-edge asymmetry for 
the KK graviton (dotted magenta) 
and the case of a spin-2 resonance mostly produced via $q\bar{q}$ (dashed green).
}
\label{fig:Campane}
\end{figure}

In fig.~\ref{fig:Campane} we show our results for the center-edge asymmetry. 
For the sake of simplicity we consider only signal events.
We compare the distributions for the four cases proposed in fig.~\ref{fig:SpinqQuark}.
For both tight (left panel) and loose (right panel) cuts 
the KK graviton distribution shifts towards the spin-0 one if compared with the results of section~\ref{sec:Results}.
The reduced distance between the two cases corresponds to a lower separation in terms of statistical significance.
Similar results can be obtained by means of the LLR.

\begin{figure}[!htb!]
\centering
\minipage{0.5\textwidth}
  \includegraphics[width=.8\linewidth]{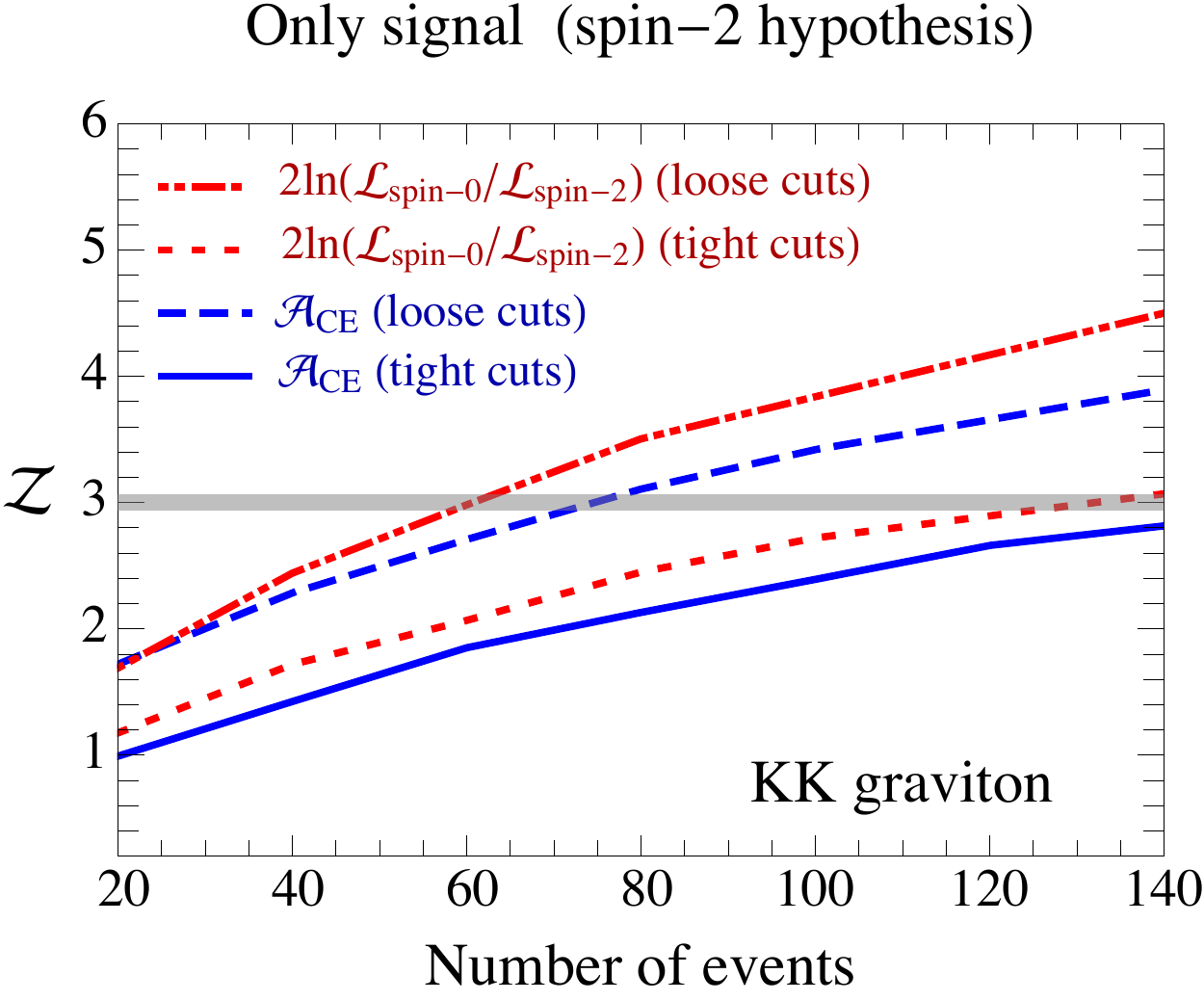}
\endminipage\hfill
\minipage{0.5\textwidth}
  \includegraphics[width=.8\linewidth]{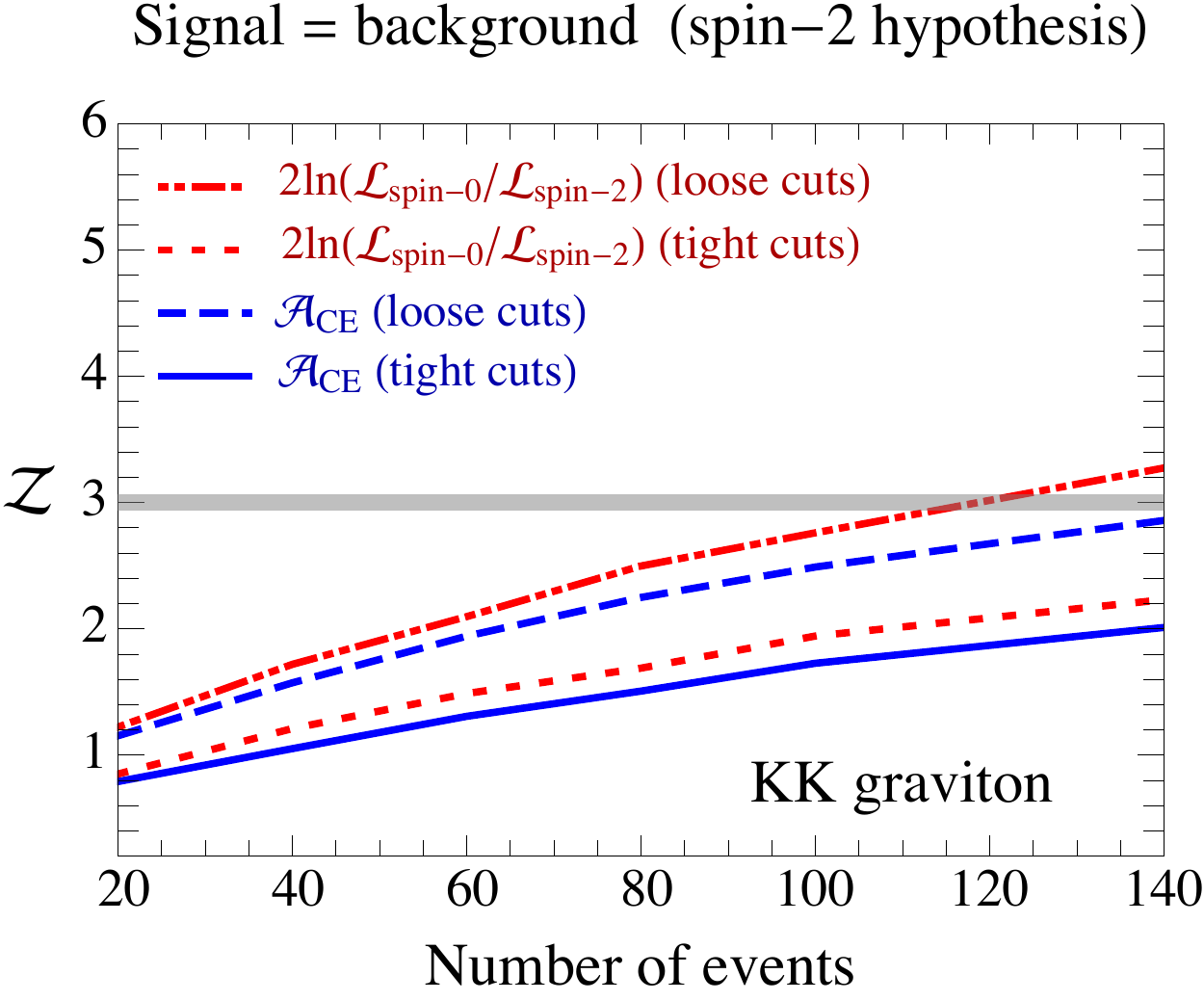}
\endminipage
\caption{\em 
Significance  $\mathcal{Z}$ for the LLR and the center-edge asymmetry as a function of the number of events.
The significance refers to the spin-2 hypothesis in the case of a KK graviton with universal couplings.
}
\label{fig:KK}
\end{figure}

The main result of this section is summarised in fig.~\ref{fig:KK}.
We show the significance $\mathcal{Z}$ as a function of the collected number of events.
As in fig.~\ref{fig:MoneyPlot} we compare the results obtained 
using the LLR and the center-edge/dartboard asymmetry (see caption and labels for details).

On the qualitative level our findings are similar to those already discussed in section~\ref{sec:Results}.
Loose cuts perform slightly better than tight ones,
and the LLR
gives better result than the center-edge asymmetry.
Considering only signal events (left panel in fig.~\ref{fig:KK}) and loose cuts
a significance of $\mathcal{Z}\simeq 3$ ($\mathcal{Z}\simeq 4$) can be reached with $N_{\rm obs}^{(J)} = 20$ ($N_{\rm obs}^{(J)} = 100$) events, and
slightly worse results---significance of $\mathcal{Z}\simeq 2$ ($\mathcal{Z}\simeq 3$) with $N_{\rm obs}^{(J)} = 20$ ($N_{\rm obs}^{(J)} = 100$) events---are possible  
by imposing tight selection cuts.
The inclusion of background events reduces the expected significance. In the extreme case in which we include
in our simulated samples an equal number  of signal and background events (right panel in fig.~\ref{fig:KK})
we expect at most a $\mathcal{Z}\simeq 3$ significance for the spin-2 KK graviton hypothesis 
with loose selection cuts and $N_{\rm obs}^{(J)} =N_{\rm obs}^{\rm (bkg)} = 100$.

The significance $\mathcal{Z}=3$ threshold is reached for 70 (140) events for the signal with loose (tight) selection cuts. The situation is therefore different from before for the case of a spin-2 resonance produced only by gluon fusion. The choice of selection cuts makes a difference and a luminosity of about 9 fb$^{-1}$ is necessary in the case of the loose cuts whereas a larger luminosity of 25 fb$^{-1}$ is required by the tight selection cuts.

\subsection{Estimating the model-dependent systematic uncertainty} \label{sec:uncertainty}


\begin{figure}[!h]
\centering
  \includegraphics[width=.4\linewidth]{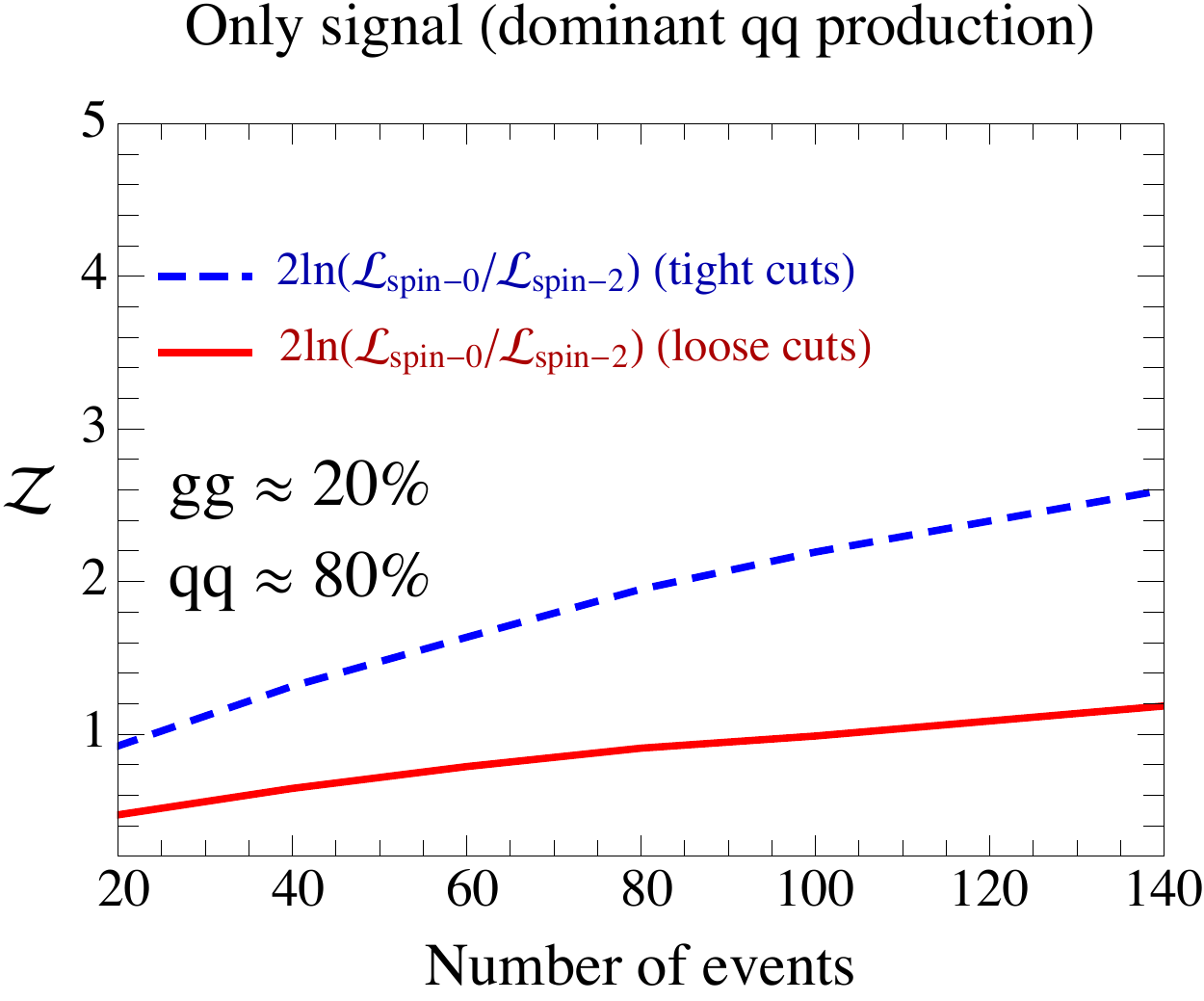}
\caption{\em 
Significance $\mathcal{Z}$ for the LLR  as a function of the number of events.  In the model for the spin-2 resonance, the production mechanism is taken to be  dominated
 by $q\bar{q}$ production.
}
\label{fig:QQcase}
\end{figure}

The presence of the $q\bar{q}$ production mechanism reduces the
discriminating power of the angular distributions.
For the case of the KK gravitons,
   we find a reduction of the significance $\mathcal{Z}$
of the order of 10\%-30\% (depending on the analysis) with respect to the case in which the spin-2 resonance is produced only via gluon fusion.

A more significant reduction takes place if the $q\bar{q}$ production mechanism becomes more important. Fig.~\ref{fig:Campane} shows the reduced discrimination power as the $q\bar{q}$ production becomes the dominant one. This reduction in significance is due to a systematic error intrinsic to  the definition of the model behind the spin-2 hypothesis  in so far as the angular distribution depends on the production mechanism. 

The case in which the $q\bar{q}$ production mechanism is dominant  requires a larger number of events in order to tell the two spin possibilities apart.  Fig.~\ref{fig:QQcase} shows the significance in the extreme case (still compatible with other experimental constraints) of 
 $q\bar{q}$  accounting for 80\% of the production.

In this case, it is not possible to reach the $\mathcal{Z}= 3$ level with a reasonable number of events and one must include in the analysis other decay channels.  For the case of the Higgs boson, it has been shown~\cite{Aad:2013xqa} that the decay channel into two $W$ gauge bosons  enjoys an angular variable that is less sensitive to the production mechanism. This channel, together with that into two $Z$ bosons, can also be used to distinguish the parity of the resonance. We postpone such an analysis to a time after the existence of the resonance has been ascertained.

Finally, one must bear in mind that many other systematic errors---for example, those on  the  integrated luminosity and selection efficiency or those on the photon  identification---are lurking around different steps of the analysis. Their impact could in principle be estimated by incorporating the uncertainties into the likelihoods or convoluting the pdf with the corresponding \textit{nuisance} parameters. We have not attempted  such a procedure here because a realistic estimate  of these uncertainties can only be done by the experimental collaborations.

\begin{acknowledgments}
We thank Roberto Franceschini, Nicola Orlando and Alberto Tonero for discussions. MF thanks the International School for Advanced Studies (SISSA) and the Physics Department of the University of Trieste for the hospitality.
\end{acknowledgments}

 \appendix
 

 \section{Comparison with the ATLAS results}
\label{app:A} 


\begin{figure}[!htb!]
\centering
\minipage{0.5\textwidth}
 \includegraphics[width=.8\linewidth]{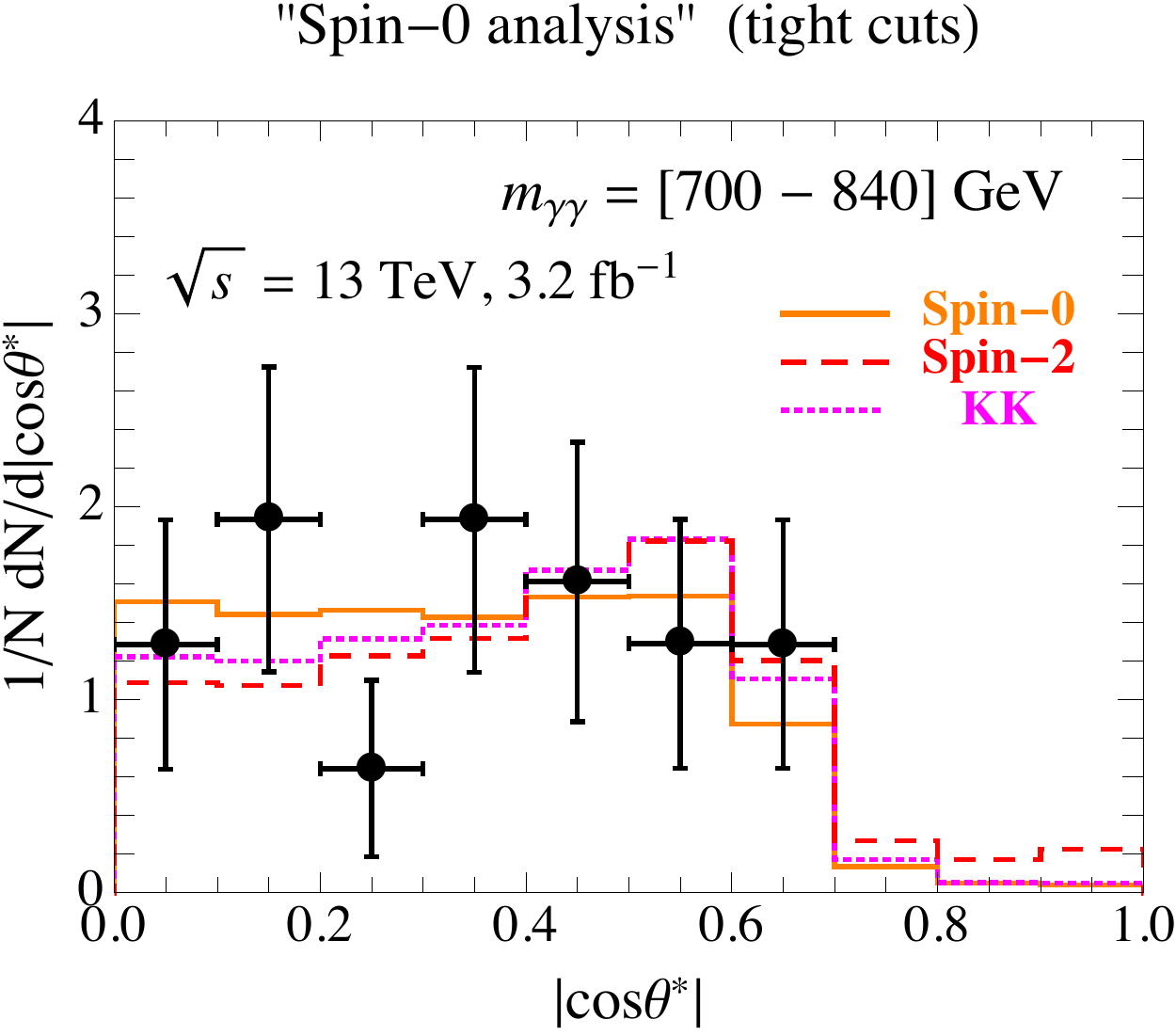}
\endminipage\hfill
\minipage{0.5\textwidth}
 \includegraphics[width=.8\linewidth]{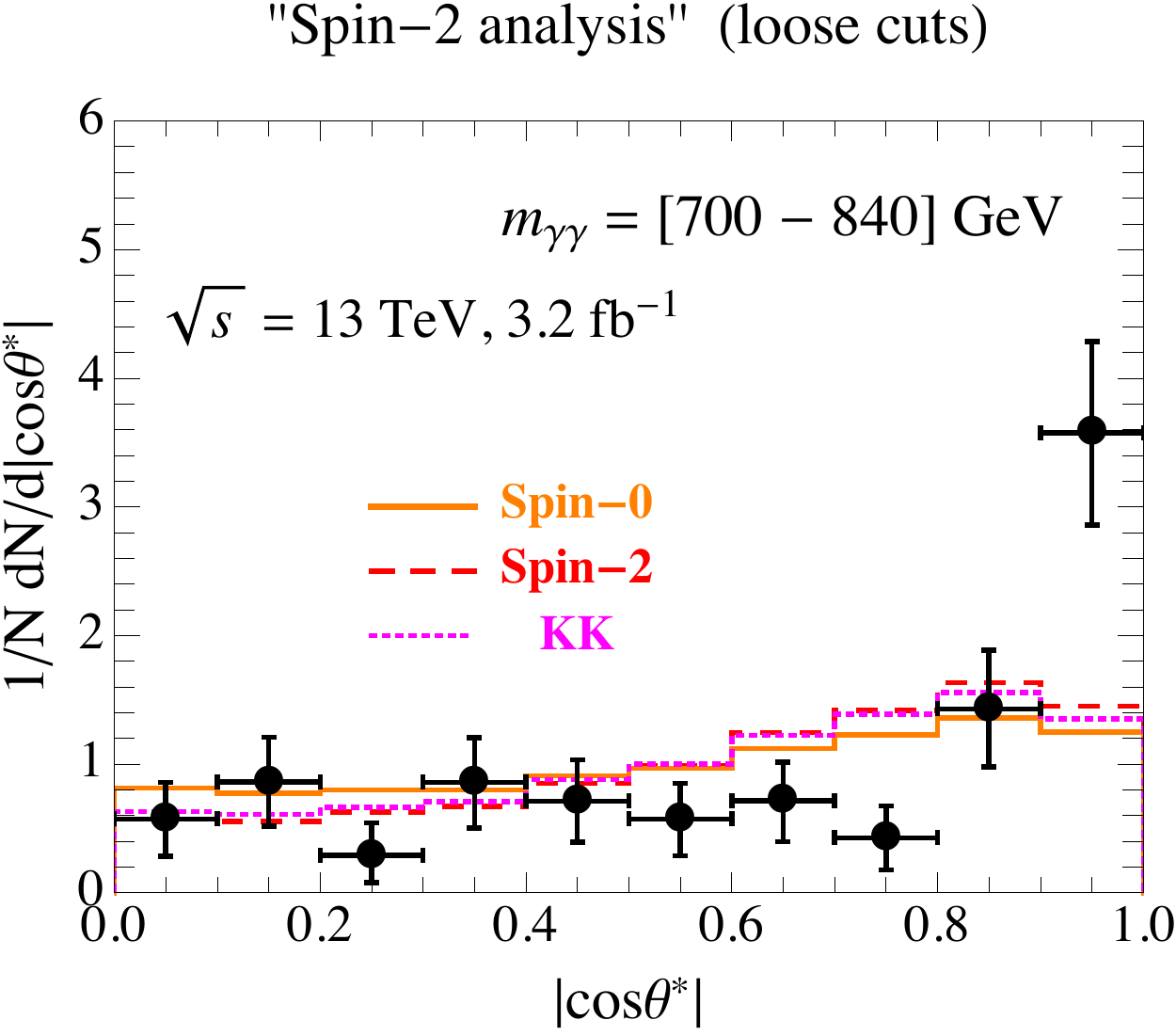}
\endminipage
\caption{\em 
Distribution of $|\cos\theta^*|$ for events in the mass interval $m_{\gamma\gamma} =[700-840]$ GeV.
Data points are digitalised from the ATLAS reference in~\cite{ATLASMoriond}.
We superimpose the distributions for the signal-plus-background events generated
in section~\ref{sec:Results} for both the spin-0 and spin-2 case.
Signal and background distributions are weighted in such a way to reproduce 
$N_{\rm obs}^{(J)} = 18$ signal and $N_{\rm obs}^{(\rm bkg)} = 12$ background events for the ``spin-0 analysis'' (left panel) 
and $N_{\rm obs}^{(J)} = 25$ signal and $N_{\rm obs}^{(\rm bkg)} = 45$ background events for the ``spin-2 analysis'' (right panel).
}
\label{fig:ATLAScomparison}
\end{figure}
The ATLAS collaboration presented two analysis: the first one---dubbed ``spin-0 analysis'' in~\cite{ATLASMoriond}, and shown in the left panel of fig.~\ref{fig:ATLAScomparison}---is optimised for a spin-0 resonance search 
while the second one---dubbed ``spin-2 analysis'' in~\cite{ATLASMoriond}, and shown in the right panel of fig.~\ref{fig:ATLAScomparison}---is optimised for a spin-2 resonance search. 
In this paper, the analysis with tight (loose) selection cuts corresponds to the ``spin-0 analysis'' (``spin-2 analysis'') in~\cite{ATLASMoriond}.
In both cases we superimpose to the
experimental data (representing the full data set, without any background subtraction) 
the pdfs for the signal-plus-background events generated
in section~\ref{sec:Results} for both the spin-0 and spin-2 case.
For the ``spin-0 analysis'' (``spin-2 analysis'') resonance search, 31 (70) data events are
observed with $m_{\gamma\gamma} = [700-840]$ GeV~\cite{ATLASMoriond}.

\begin{figure}[!h]
\centering
\minipage{0.5\textwidth}
  \includegraphics[width=.8\linewidth]{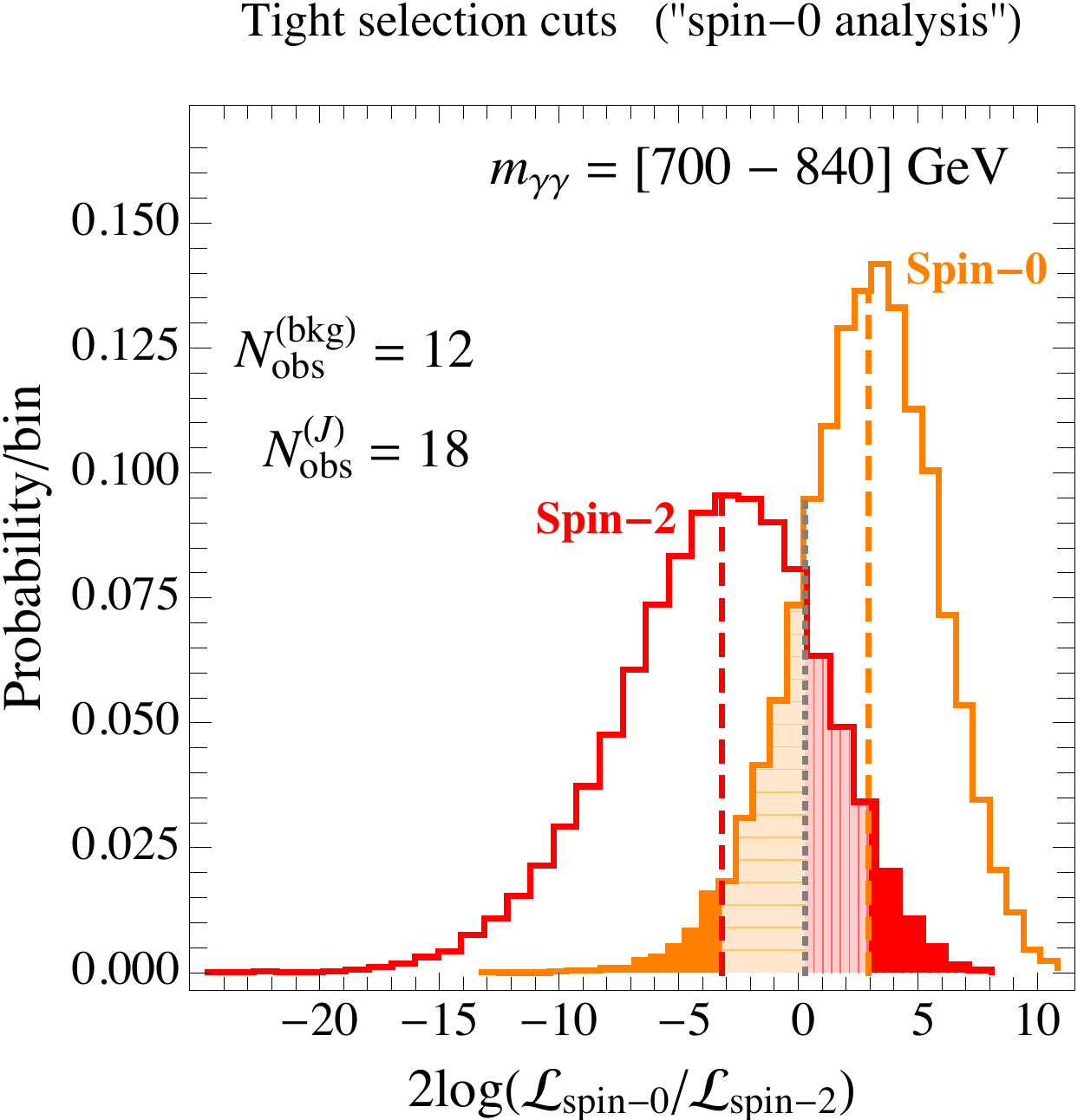}
\endminipage\hfill
\minipage{0.5\textwidth}
  \includegraphics[width=.8\linewidth]{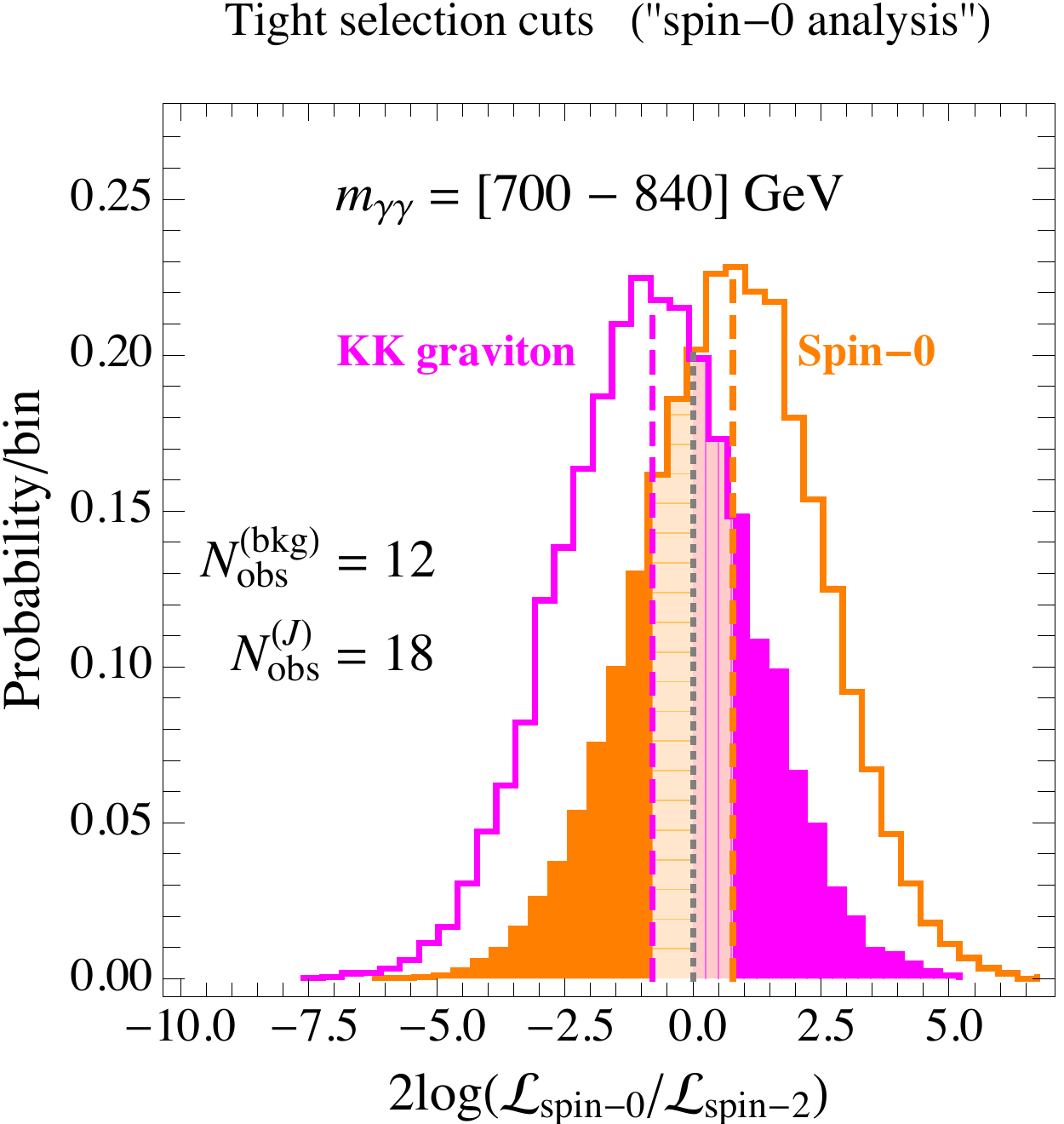}
\endminipage
\caption{\em 
LLR analysis for $N_{\rm obs}^{(J)} = 18$ signal and $N_{\rm obs}^{(\rm bkg)} = 12$ background events. 
In the left (right) panel we compare the spin-2 produced via gluon fusion (KK graviton) and spin-0 hypotheses.
The vertical dashed lines represent the medians of the two distributions (see also caption of fig.~\ref{fig:ACESignal}). 
}
\label{fig:ATLASCampane}
\end{figure}

Let us start considering the ``spin-0 analysis''.  The fit of the mass-invariant spectrum is shown in fig.~\ref{fig:Fit}, and from our best-fit values
 in 
table~\ref{tab:Fit} we find $N_{\rm obs}^{(J)} = 18.0$ signal and $N_{\rm obs}^{(\rm bkg)} = 12.4$ background events---in agreement with the total value
of 31 events quoted by the ATLAS collaboration.
We therefore apply the LLR analysis outlined in section~\ref{sec:LLRanalysis}, 
simulating $N_{ps} = 10^4$ pseudo-experiments with $N_{\rm obs}^{(J)} = 18$ signal and $N_{\rm obs}^{(\rm bkg)} = 12$ background events.
We use tight selection cuts since they are equivalent---as stated above---to the ``spin-0 analysis''.
We show our results in fig.~\ref{fig:ATLASCampane} 
where we compare the spin-0 hypothesis with  both  the spin-2 produced via gluon fusion and the KK graviton (respectively, in the left and right panel).

Even though it is not possible to derive any definite conclusions from these very preliminary results,
it is interesting to try an hypothesis testing.
Assuming the spin-0 nature of the resonance---that is sitting on the median of the corresponding distribution, see fig.~\ref{fig:ATLASCampane}---we can estimate the statistical significance of this choice by computing the probability to accept the spin-0 hypothesis when it is wrong (that is, an error of Type I represented by the red and magenta regions
in  fig.~\ref{fig:ATLASCampane}). Alternatively, one can do the opposite assuming the spin-2 nature of the resonance.
As far as the case with the KK graviton is concerned, as clear from the right panel of fig.~\ref{fig:ATLASCampane} 
it is not possible to derive any conclusion since the two hypothesis are equally preferred by the LLR analysis.
Considering the left panel of  fig.~\ref{fig:ATLASCampane}, on the contrary,
we find the $p$-value $P_2 \simeq 0.048$ for the spin-0 hypothesis, corresponding to the statistical significance $\mathcal{Z} \simeq 1.66$.
Assuming the spin-2 hypothesis, we find the $p$-value $P_0 \simeq 0.028$, corresponding to the statistical significance $\mathcal{Z} \simeq 1.89$.
We find similar results taking into account the ``spin-2 analysis'' with $N_{\rm obs}^{(J)} = 25$ signal and $N_{\rm obs}^{(\rm bkg)} = 45$ background events (we do not show the
corresponding distributions, qualitatively equivalent to the one shown in fig.~\ref{fig:ATLASCampane}).
Considering the case with the spin-2 resonance produced via gluon fusion, 
we find the $p$-value $P_2 \simeq 0.047$ for the spin-0 hypothesis, corresponding to the statistical significance $\mathcal{Z} \simeq 1.67$.
Assuming the spin-2 hypothesis, we find the  $p$-value $P_0 \simeq 0.033$, corresponding to the statistical significance $\mathcal{Z} \simeq 1.83$.

Finally, by looking at the result of the ``spin-2 analysis'' in the right panel of fig.~\ref{fig:ATLAScomparison},  
the presence of a discrepancy in the last few bins catches the eye.
%
%
As far as the analysis with loose selection cuts (or, equivalently, the ``spin-2 analysis'' in~\cite{ATLASMoriond}) is concerned,
we do not find any particular peak in the forward region with $|\cos\theta^*| > 0.8$ even in the presence of a spin-2 signal compatible in strength with the 
observed excess. This is expected, since the forward region 
is depleted by limited detector acceptances. Of course one may argue that our phenomenological analysis  
cannot reproduce the same accuracy in photon 
identification and isolation reached by the experimental analysis; despite being this objection very true, the presence of 
similar discrepancies in the forward region also in the sidebands with $m_{\gamma\gamma} = [600-700]$ GeV and $m_{\gamma\gamma} = [840-\infty]$ GeV (see the corresponding 
plots 
in~\cite{ATLASMoriond}) seems to point towards the existence of some uncontrolled systematics. 

Bearing all  this in mind, 
let us entertain the possibility that the discrepancy in the forward region is actually due to the presence of a spin-2 signal.
It follows that this effect, if real, should be 
related to something that is not captured by the simulated signal-plus-background events.
One intriguing possibility is represented by non-trivial interference effects between 
the spin-2 signal and the SM background. We discuss  this case in the next sub-section.


\subsection{Resonance-continuum interference for a spin-2 state}

The interference between resonance and continuum---that is the interference between diagrams $D$ and $E$ in fig.~\ref{fig:Diagrams}---cannot be neglected   in the di-photon Higgs signal at the LHC, as pointed out in~\cite{Dixon:2003yb}.
In~\cite{Jung:2015etr} the analysis was extended for the di-photon excess at 750 GeV but only in the case of a scalar resonance. 
Let us focus on the relevant points of the computation. 
We consider production via gluon fusion of a resonance with mass $M_X$ and
width $\Gamma_X$.
At the parton level with center of mass $\hat{s}$ the amplitude is
\begin{equation}
\mathcal{A}_{gg\to \gamma\gamma} = -\frac{\mathcal{A}_{gg\to X} \mathcal{A}_{X\to \gamma\gamma}}{\hat{s} - M_X^2 + iM_X \Gamma_X} + \mathcal{A}_{\rm cont}~,
\end{equation}
where $\mathcal{A}_{\rm cont}$ is the amplitude for the background process generated by gluon fusion
while $\mathcal{A}_{gg\to X}$ and $ \mathcal{A}_{X\to \gamma\gamma}$ are the amplitude for the production of $X$ and the subsequent di-photon decay.
In the SM the amplitude $\mathcal{A}_{\rm cont}$  arises at one-loop level (see fig.~\ref{fig:Diagrams}).
The interference term is~\cite{Dixon:2003yb} 
\begin{equation}\label{eq:Interference}
\delta\hat{\sigma}_{gg\to X\to \gamma\gamma} = -2\left(\hat{s} - M_X^2\right)
\frac{\Re\left(\mathcal{A}_{gg\to X} \mathcal{A}_{X\to \gamma\gamma} \mathcal{A}^{*}_{\rm cont}\right)}{\left(\hat{s} - M_X^2 \right)^2 + M_X^2 \Gamma_X^2}
-2M_X\Gamma_X \frac{\Im\left(
\mathcal{A}_{gg\to X} \mathcal{A}_{X\to \gamma\gamma} \mathcal{A}^{*}_{\rm cont}
\right)}{\left(\hat{s} - M_X^2 \right)^2 + M_X^2 \Gamma_X^2}~,
\end{equation}
and the corresponding contribution at the hadron level is
\begin{equation}
\delta\sigma_{pp\to X\to \gamma\gamma} = \int\frac{d\hat{s}}{\hat{s}} \frac{dL_{gg}}{d\hat{s}}\delta\hat{\sigma}_{gg\to X\to \gamma\gamma}~,
\end{equation}
with $dL_{gg}/d\hat{s}$ the gluon luminosity function.
We start briefly discussing the case of the Higgs boson.
The first term in eq.~(\ref{eq:Interference}) turns out to be zero in the narrow-width approximation 
when integrated over an invariant-mass bin centered on the resonance. 
The second term in eq.~(\ref{eq:Interference}) needs a large imaginary part to give a sizeably contribution.
For the Higgs, the largest imaginary  contribution arises at two-loop level from the background $gg\to \gamma\gamma$ amplitude (at one loop  
the spin-0 nature of the Higgs boson selects, as a consequence of helicity conservation, 
only the like-helicity states $g^{\pm}g^{\pm}$ and $\gamma^{\pm}\gamma^{\pm}$ whose amplitudes are suppressed by the factor $m_q^2/m_H^2$~\cite{Dicus:1987fk}).
Despite the two-loop suppression, for a Higgs boson the resonance-continuum interference generates 
non-negligible effects in particular 
in the forward direction.

\begin{figure}[!htb!]
\centering
\minipage{0.5\textwidth}
 \includegraphics[width=.8\linewidth]{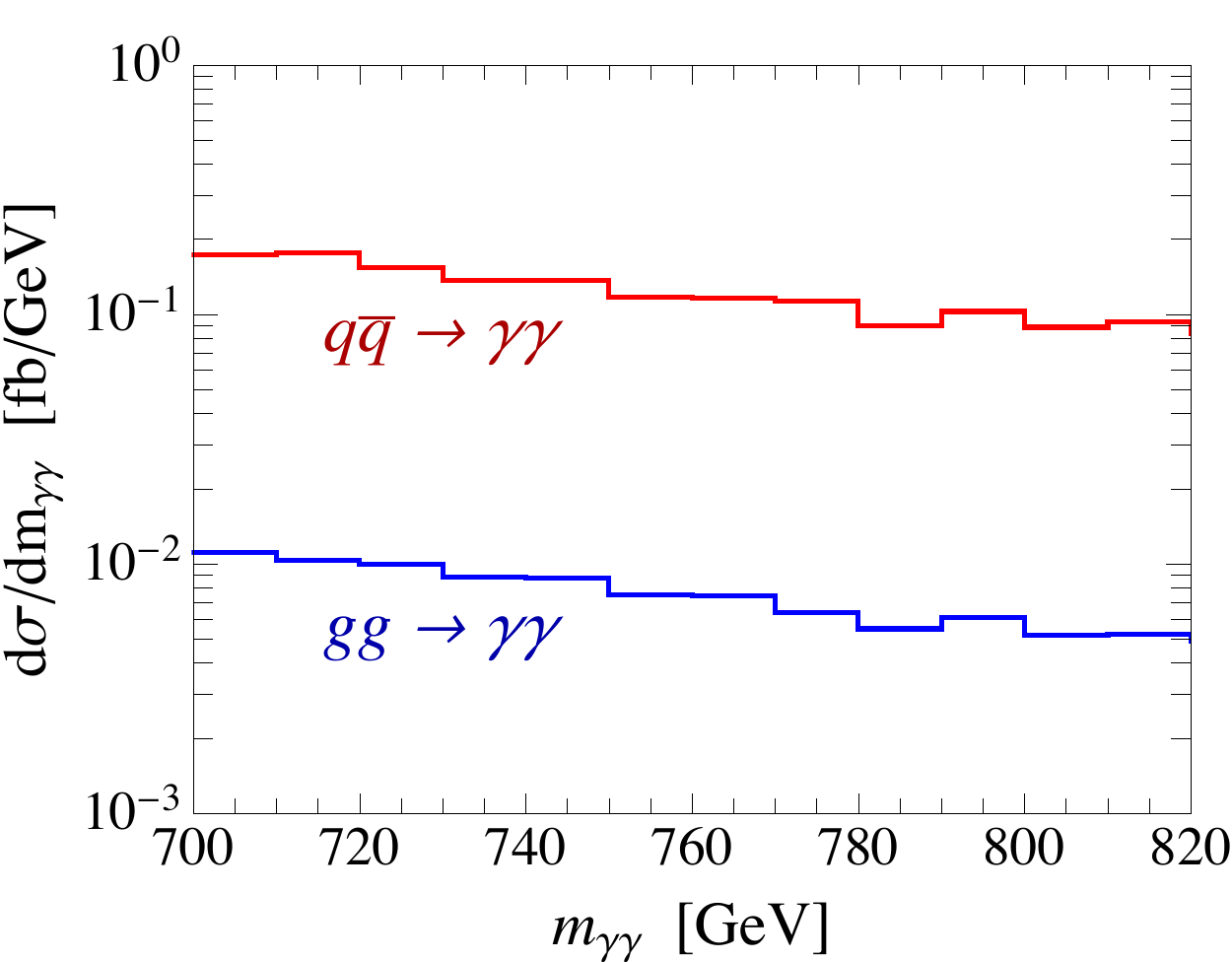}
\endminipage\hfill
\minipage{0.5\textwidth}
 \includegraphics[width=.8\linewidth]{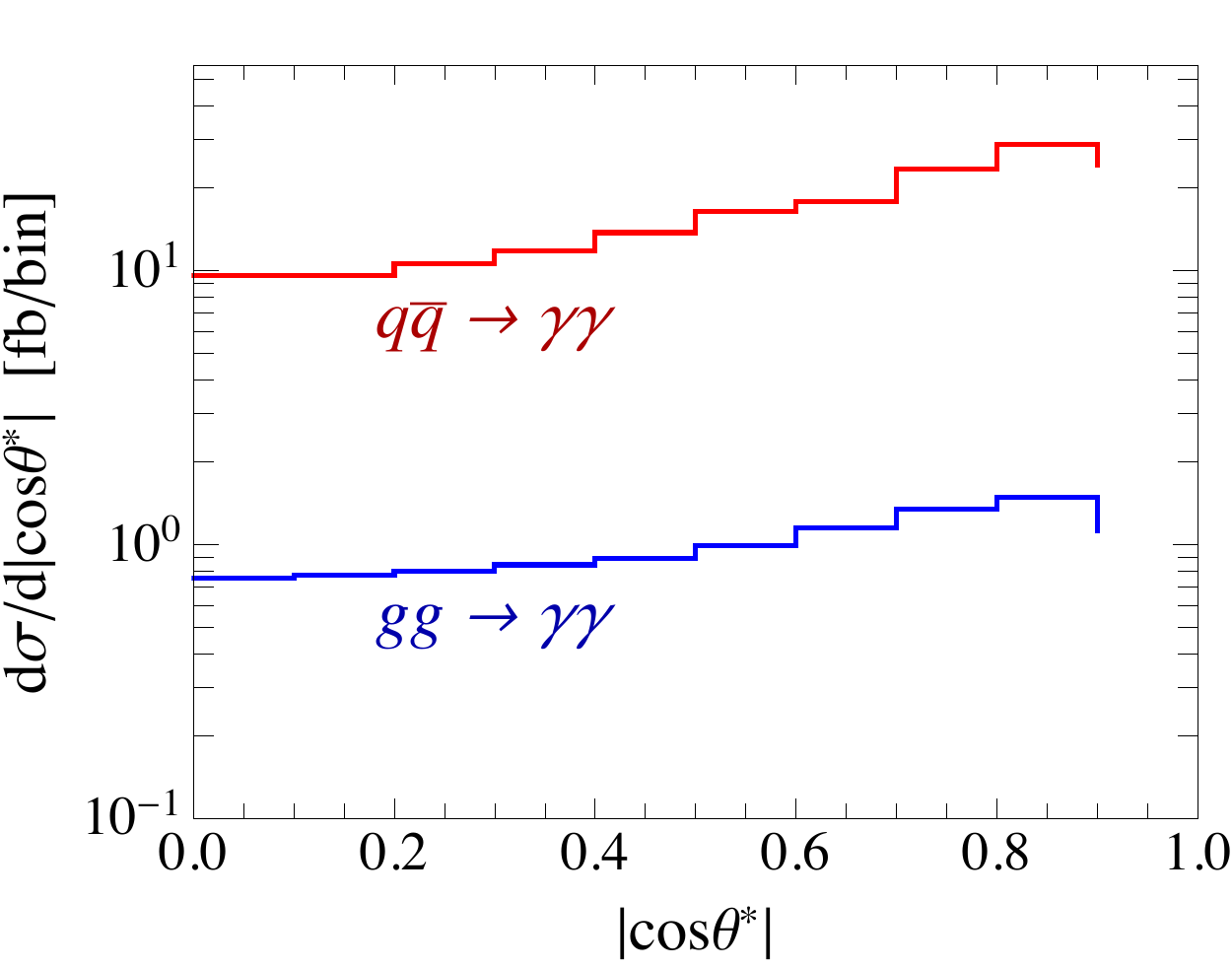}
\endminipage
\caption{\em 
Comparison between the tree level cross-section $q\bar{q} \to \gamma\gamma$ (red) and the one loop 
process $gg \to  \gamma\gamma$ (blue) at the LHC with $\sqrt{s} = 13$ TeV.
We show in the left panel the differential cross-section $d\sigma/dm_{\gamma\gamma}$ as a function of the di-photon invariant mass, 
and in the right panel  the differential cross-section $d\sigma/d|cos\theta^*|$ as a function of the 
scattering angle in the CS frame (see fig.~\ref{fig:CSFrame}). 
In the right panel we include only events with $m_{\gamma\gamma}  = [700-840]$ GeV.
}
\label{fig:OneLoop}
\end{figure}

Motivated by this observation, it would be interesting---if the di-photon excess will be confirmed in the near future with a persisting discrepancy in the 
forward direction---to generalise the analysis to the case of a spin-2 resonance.
In the following we only list some differences with respect to the case of the spin-0 Higgs:

\begin{itemize} 
\item If the hint in favor of a broad resonance ($\Gamma_X \sim 6\%\,M_X$) will be confirmed, 
the narrow-width approximation is no longer applicable to the first term in eq.~(\ref{eq:Interference}).
As discussed in~\cite{Jung:2015etr}, the net effect of this term is a distortion of the shape of the resonance.
\item The term proportional to the imaginary part in eq.~(\ref{eq:Interference}) is proportional to the width $\Gamma_X$, and it is enhanced in the large-width scenario.
Furthermore, contrary to the Higgs case, a large imaginary part in $\mathcal{A}_{gg\to X} \mathcal{A}_{X\to \gamma\gamma} \mathcal{A}^{*}_{\rm cont}$
may be generated in  $\mathcal{A}_{gg\to X}$ and  $\mathcal{A}_{X\to \gamma\gamma}$ 
if $M_X > 2M_{V}$, where $M_{V}$ generically  denotes the mass of the particles---either SM, like the $W$ or the top, or new vector-like states---running in the loop.
\item As far as the imaginary part coming from $\mathcal{A}^{*}_{\rm cont}$ is concerned, 
as mentioned before
in the case of a spin-0 resonance one is forced to consider only amplitudes with like-helicity states $g^{\pm}g^{\pm}$ and $\gamma^{\pm}\gamma^{\pm}$ that 
are purely real.
In the case of a spin-2 particle, on the contrary, this selection rule is no longer valid since all possible helicity combinations are allowed.
One may therefore get a large contribution already at one loop by overcoming the mass suppression $m_q^2/m_H^2$ characteristic 
of the scalar Higgs case. 

The helicity structure of the one-loop process $gg \to \gamma\gamma$ was studied in~\cite{Dicus:1987fk}.
There are $16$ helicity structures $\mathcal{A}_{\lambda_1 \lambda_2 \to \lambda_3 \lambda_4}(s,t,u)$
contributing to $\mathcal{A}_{\rm cont}$, 
where $\lambda_{1,2}$ and $\lambda_{3,4}$ denote, respectively,  the helicities of the incoming gluons and outgoing photons, and $s$, $t$, $u$ are the usual Mandelstam variables for the scattering process $gg \to \gamma\gamma$.
Only three of them are really needed to 
reconstruct the full amplitude, since all the remaining ones can be derived 
using crossing relations, parity and permutation symmetry. 

Following~\cite{Dicus:1987fk}, we focus on 
$\mathcal{A}_{++\to ++}(s,t,u)$, $\mathcal{A}_{++\to +-}(s,t,u)$, and $\mathcal{A}_{++\to --}(s,t,u)$. 
In the massless limit for the quarks running in the box diagram,
the only imaginary part with like-helicity states---relevant for the interference with a spin-0 resonance---is~\cite{Dicus:1987fk}
\begin{equation}
\Im[\mathcal{A}_{++\to ++}(s,t,u)] = 
-\pi [\vartheta(t) - \vartheta(u)]\times
\left[
\frac{t-u}{s} + \frac{t^2 + u^2}{s^2} \log\left|\frac{t}{u}\right|
\right]~,
\end{equation}
that however vanishes since $t = -(s/2)(1-\cos\theta) < 0$, $u = -(s/2)(1+ \cos\theta) < 0$, with $\theta$ the scattering angle in the center of mass frame.

In the presence of a spin-2 resonance, 
the interference involves amplitudes with 
different helicities in the initial and final state. Using crossing symmetry, we find
\begin{eqnarray}
\Im[\mathcal{A}_{+-\to +-}(s,t,u)]  &=&   -\pi [\vartheta(t) - \vartheta(s)]\times
\left[
\frac{t-s}{u} + \frac{t^2 + s^2}{u^2}\log
\left|
\frac{t}{s}
\right|
\right]~,\label{eq:Amplitude1}
\\
\Im[\mathcal{A}_{+-\to -+}(s,t,u)]  &=& -\pi [\vartheta(s) - \vartheta(u)]\times
\left[
\frac{s-u}{t} + \frac{s^2 + u^2}{t^2} \log\left|\frac{s}{u}\right|
\right]~.,\label{eq:Amplitude2}
\end{eqnarray}
This very simple computation 
shows that in the presence of a spin-2 resonance
large imaginary contributions from the continuum part in the second term of eq.~(\ref{eq:Interference}) are possible: 
they are described by the amplitudes in eqs.~(\ref{eq:Amplitude1},\,\ref{eq:Amplitude2}), and they are not suppressed by the quark masses. 
\item Finally, it is important to keep in mind another important difference with respect  to the Higgs case. 
Let us consider the various diagrams participating in the definition of the irreducible SM background, depicted in fig.~\ref{fig:Diagrams}.
In the invariant mass range $m_{\gamma\gamma} = [100-200]$ GeV---relevant for di-photon Higgs searches--- the 
one-loop amplitude $gg\to\gamma\gamma$ is comparable in size with 
the tree-level non-resonant di-photon process $q\bar{q} \to \gamma\gamma$. 
This is no longer true in the mass range $m_{\gamma\gamma} = [700-840]$ GeV
where the one-loop amplitude $gg\to\gamma\gamma$ turns out to be 
one order of magnitude smaller than the tree-level process $q\bar{q} \to \gamma\gamma$, as shown in fig.~\ref{fig:OneLoop}.
As a result,  resonant-continuum interference involving the one-loop amplitude $gg\to\gamma\gamma$
has to overcome this large suppression in order to give a sizeable correction to the di-photon signal rate.

\end{itemize}

\clearpage

\end{document}